\documentclass[prb,eqsecnum,showpacs,byrevtex,amsmath,amssymb,nofootinbib]{revtex4}
\usepackage{amsmath,graphicx,psfrag}
\begin{document}
   
\title{Globular Structures of a Helix-Coil Copolymer: \\
Self-Consistent Treatment.}
\author{C. Nowak, V.G. Rostiashvili, and T.A. Vilgis}
\affiliation{Max-Planck-Institut f\"ur Polymerforschung, Ackermannweg 10 , 55128 Mainz, Germany}
\begin{abstract}
A self-consistent field theory was developed in the grand-canonical ensemble formulation to study transitions in a helix-coil multiblock globule. Helical and coil parts are treated as stiff rods and self-avoiding walks of variable lengths correspondingly. The resulting field-theory takes, in addition to the conventional Zimm-Bragg parameters, also three-dimensional interaction terms into account. The appropriate differential equations  which determine the self-consistent fields were solved numerically with finite element method. Three different phase states are found: open chain, amorphous globule and nematic liquid-crystalline (LC) globule. The LC-globule formation is driven by the interplay between the hydrophobic helical segments attraction and the anisotropic globule surface energy of an entropic nature. The full phase diagram of the helix-coil copolymer was calculated  and thoroughly discussed. The suggested theory shows a clear interplay between secondary and tertiary structures in globular homopolypeptides.
\end{abstract}
\maketitle
\section{Introduction}
Significant work has been done to investigate the helix-coil transition theoretically~\cite{Poland} 
and computationally~\cite{smith,ireta}. One of the first 
and  well-known approaches is the Zimm-Bragg theory~\cite{zimm} which is designed for single-stranded polypeptide chains. It considers a one-dimensional Ising type model 
in which each segment can be in one of the two states: helical state (stiff) or coil state (flexible). The helix is stabilized 
by hydrogen bonds which generates an energy gain of $-\epsilon$ for each segment in the helical state. This energy gain is 
partly balanced by an entropy loss $-\Delta S$. The free energy difference between the helical state and the coil state is 
therefore given by $\Delta f = -\epsilon + T \Delta S$ for each segment. 

In an $\alpha$-helix hydrogen bonds can be only formed  between 
the $i^{\rm th}$ and the $(i+3)^{\rm th}$ peptide group.  The formation of a hydrogen bond between 
the first and the third peptide group fixes the conformation of three groups. The next bond between the second and the fourth group furnishes the same energy gain but fixes only one new group and thus leads to a much smaller entropy loss. The formation of an $\alpha$-helix ( as a simple  element of the secondary structure) is therefore a cooperative process and the formation of a junction between helix and coil is energetically unfavorable.  This can be modelled by an energy penalty $\mu_{\rm J}$ for each junction 
between two successive groups  of helical  and  flexible segments. Similar arguments hold for all kind of helices, such that  the helix formation is always a cooperative process. 

It is convenient to define 
the following fugacities 
\begin{equation}
s \equiv e^{-\beta \Delta f}, \quad \sigma \equiv e^{-2\beta \mu_{\rm J}},
\end{equation}
where $s$ gives the statistical weight of a helical segment compared to a coil segment. 
The cooperativity parameter $\sigma$  gives the statistical weight of a junction point. 
$\sigma=1$ corresponds to $\mu_{\rm J}=0$ and therefore to a non-cooperative system. 
$\sigma \rightarrow 0$ corresponds to $\mu_{\rm J} \rightarrow \infty$ and therefore to a totally cooperative system, i.e. either the 
entire chain forms one big helix or no helix is formed at all. In most helix forming biopolymers $\sigma$ is roughly $10^{-3}-10^{-4}$ (see e.g. ref.~\cite{Poland}).

Using a transfer matrix method~\cite{Poland} the one dimensional model can be solved exactly.  The $s$ - dependance of  the  fraction of stiff helical segment $\Theta_{\rm R} = N_{\rm R}/N$   has a typical sigmoid - type  form .  With increasing cooperativity (i.e. as the parameter $\sigma$ is reduced)  the transition 
becomes sharper and sharper.  
This rather sharp crossover transition (due to the cooperativity  effect) is also observed experimentally, for instance in polybenzylglutamate~\cite{zimm}.

Several extensions have been made to this one-dimensional Ising type model. 
It has been shown~\cite{farago} that the transition becomes less cooperative, if  the hydrogen-bonding ability of the solvent  is taken into account. The helix-coil transition in grafted chains was studied~\cite{buhot} as well as the effect of an external applied force on the transition~\cite{mario}. 

Of special interest is the application of one-dimensional models to proteins, see for instance~\cite{tanaka,coolen}.  A study of the helix-coil transition including long-range electrostatic interactions~\cite{vasquez} can, to some extent, explain the amount and location of helical segments in globular proteins. However, to understand how $\alpha$-helices (or generally 
secondary structure elements) are formed in the folding process of a protein and how this influences the compaction and formation of tertiary structure (and vice versa), it is necessary 
to combine the one-dimensional physics of the helix-coil transition with the  interactions of segments which approach each other in the three-dimensional space. This enables a description of the interplay between the {\it secondary structure}  and the mesoscopic structure formation of the entire chain (known as {\it tertiary structure}) . 
   
 In proteins the stiff helical parts are often hydrophobic since the hydrogen bonds stabilizing the helix composition (as opposed to coil parts)  are mainly saturated (due to bond formation between the $i^{th}$ and the $(i+3)^{th}$ peptide groups).  This hydrophobicity causes an additional attractive interaction between stiff parts and drives the protein into  a compact globular phase.  Statistical analysis of the data from 41 globular proteins in native and partially folded conformational states~\cite{uversky} showed a strong correlation between the amount of secondary structure elements and compactness of the proteins. This indicates that the formation of secondary structure (for instance $\alpha$-helices) and the hydrophobic collapse into a compact globule occur simultaneously.  This problem has been partially discussed within computer simulations of globular proteins~\cite{Dill,socci,Shakhnovich,sikorski}. Among other things, it has been shown \cite{socci,Shakhnovich,sikorski} (as opposed to the earlier findings by Dill et al.\cite{Dill}) that the compactness itself, driven by the hydrophobicity, is insufficient to generate any appreciable secondary structure. It was necessary to introduce a local conformational propensity toward $\alpha$ - helix formation. The interplay of compaction and secondary structure  leads to the formation of the specific three-dimensional tertiary structure. Computational and experimental studies of this mechanism can for instance be found in ref.~\cite{clementi,mayor}. 

To study the interplay of helix-coil (or stiff-flexible) transition and collapse transition of the polymer into a compact globule, we have  developed an approach which combines variable composition with three-dimensional excluded volume interactions using self-consistent field theory.
The paper is organized as follows. In Section II we have covered the general self - consistent field theory of a  helix - coil copolymeric globule. The final equations are written down  in a form which is convenient for the numerical solution.  Section III is devoted to the analysis of this  solution. The formation of different globule structures (e.g. amorphous and liquid - crystalline (LC) globules) are studied in details. Among other things we argue that the LC - globule formation is mainly driven by the globule surface tension anisotropy. Finally,  in Section IV we sketch the main results and compare them with the appropriate findings of some other authors.
\section{Theory}
\begin{figure}[h!]
\begin{center}
\includegraphics[width=0.35\linewidth]{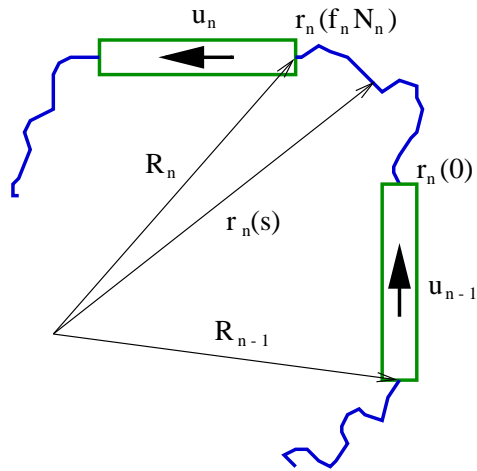}
\caption[Rod-coil multiblock copolymer parameterization]{Rod-coil multiblock copolymer 
parameterization. Two successive rod-coil blocks with ordinal numbers $n-1$ and $n$ are shown. 
The flexible chain coordinates  use the vector set ${\bf r}(s)$. The orientation of the $n^{\mathrm{th}}$ rod is denoted as 
${\bf u}_n$ and the junction point between rod and coil is  given by ${\bf R}_n$.}
\label{Block}
\end{center}
\end{figure}

In this section the derivation of the field theory is outlined. For a more detailed description see \cite{Nowak2}. 
Regarding all three dimensional interactions and entropic contributions the helices are modelled as stiff rods. Hence for 
one specific microscopic composition of helical and flexible parts the system looks like a rod-coil multiblock copolymer as shown in 
Fig.(\ref{Block}).
The multiblock copolymer may be 
composed of $K$ rod-coil blocks. The conformation of the $n^{\mathrm{th}}$rod-coil block
is given by the vector-function ${\bf r}_n(s)$ describing the
contour of the coil, by the vector ${\bf R}_n$ which gives the position of the
junction point between rod and coil and the unit vector ${\bf u}_n$ describes the orientation of
the rod, see Fig.(\ref{Block}).
The length of the $n^{\mathrm{th}}$ rod-coil block in units of the segment length $b$ is given by $N_n$. 
The fraction of the flexible segments (coil) is given by $f_n$ and the fraction of stiff segments (rod) thus by $1-f_n$.  
The microscopic flexible segment density ${\hat\rho}_{\mathrm C}({\bf r})$ and  
stiff segment density ${\hat\rho}_{\mathrm R}({\bf r})$ 
can now be defined as follows.
\begin{eqnarray}
{\hat\rho}_{\mathrm C}({\bf r}) = \sum_{n = 0}^{K} \int\limits_{0}^{f_n N_n} ds\,
\delta ({\bf r} - {\bf r}_n(s)) ,
\label{C_operator}
\end{eqnarray}
\begin{eqnarray}
{\hat\rho}_{\rm R}({\bf r}) = \sum_{n = 0}^{K} \int\limits_{0}^{(1 - f_n) N_n} ds\,
\delta ({\bf r} - {\bf R}_n - {\bf u}_n s b) .
\label{R_operator}
\end{eqnarray}
In addition an orientation density ${\hat S}^{i j}({\bf r})$ is introduced which is sensitive to the collective orientation of the system
\begin{eqnarray}
{\hat S}^{i j}({\bf r}) = \sum_{n = 0}^{K} \int\limits_{0}^{(1 - f_n) N_n} ds\,
\delta ({\bf r} - {\bf R}_n - {\bf u}_n s b) \left[u^i u^j -
  \frac{1}{3}\delta^{ij}\right].
\label{pi_operator}
\end{eqnarray}
The interaction Hamiltonian of the model can thus be written as 
\begin{eqnarray}   
\beta H_{\rm int}\left[{\hat\rho}_{\rm C}, {\hat\rho}_{\rm R}\right] &=&  \chi \int d^3r \,{\hat\rho}_{R}({\bf r}){\hat\rho}_{R}({\bf r}) + \frac{v}{2} \int d^3r \left[{\hat\rho}_{C}({\bf r}) + {\hat\rho}_{R}({\bf r})\right]^2 
\nonumber\\
&+& \frac{w}{3!} \int d^3r \left[{\hat\rho}_{C}({\bf r}) + {\hat\rho}_{R}({\bf r})\right]^3
+ g \int d^3r\, {\rm Tr}\left[{\hat S}^{i j}({\bf r}){\hat S}^{i j}({\bf r})\right],
\label{Hint}
\end{eqnarray}
where $v$ and $w$ control the strength of the global two- and three-body interactions. The parameter  $\chi$ controls a selective two-body interaction between 
the stiff segments which is caused by the hydrophobic nature of the helical parts of the chain (cf. Introduction). The last term is a alignment interaction between the rods of the standard Maier-Saupe form \cite{maier-saupe}.
 The representation of this type has a wide use in the polymeric liquid-crystal context \cite{Fredrickson}.

The canonical partition function of the entire system can be written as the functional integral over the collective coil, rod and orientation densities, $\rho_\mathrm{C}(\mathbf{r})$, 
$\rho_\mathrm{R}(\mathbf{r})$ and $S^{ij}(\mathbf{r})$ respectively, i.e.\begin{eqnarray}
Z (\{N_n\}, K) &=&  \int \mathcal{D}\rho_\mathrm{C}({\bf r}) \int \mathcal{D}\rho_\mathrm{R}({\bf r})\int \mathcal{D}
S^{i j}({\bf r}) \int \prod_{n=1}^{K} D{\bf r}_n(s)\, d^3R_n\, d^2 u_n
\nonumber\\
&\times& \delta (\rho_\mathrm{C}({\bf r}) - {\hat\rho}_\mathrm{C}({\bf
  r}))\delta (\rho_\mathrm{R}({\bf r}) - {\hat\rho}_\mathrm{R}({\bf r}))\delta
(S^{i j}({\bf r})  - {\hat S}^{i j}({\bf r}))\nonumber\\
&\times&  \delta (|{\bf u}_n| - 1) \delta({\bf r}_n(f_nN_n) - {\bf R}_n)\exp\left\{-
  \frac{3}{2b^2}\int\limits_{0}^{f_n N_n} d s \left(\frac{\partial{\bf
        r}_n}{\partial s}\right)^2 \right\}\nonumber\\
&\times& \exp \left\{-\chi \int d^3r \, \rho_\mathrm{R}({\bf r})\rho_\mathrm{R}({\bf
  r}) - \frac{v}{2}\int d^3r \left[\rho_\mathrm{C}({\bf r}) + \rho_\mathrm{R}({\bf r})\right]^2\right. 
\nonumber\\
&-& \left.\frac{w}{3!}\int d^3r \left[\rho_\mathrm{C}({\bf r}) + \rho_\mathrm{R}({\bf r})\right]^3
- g \int d^3r \, {\rm Tr}\left[S^{i j}({\bf r})S^{i j}({\bf r})\right]\right\} \quad.
\label{Partition}
\end{eqnarray} 
By making use of the integral representation for the $\delta$ - function  in Eq.(\ref{Partition}) (which results in an appearance of external fields $h_\mathrm{C}({\bf r})$, $h_\mathrm{R}({\bf r})$ and $h_S^{i j}({\bf r})$) the partition function of the entire system reads
\begin{eqnarray}
Z (\{N_n\}, K) &=& \int \mathcal{D}\rho_\mathrm{C}({\bf r}) \int \mathcal{D}\rho_\mathrm{R}({\bf r})\int
D S^{i j}({\bf r}) \int \mathcal{D} h_\mathrm{C}({\bf r})
\int \mathcal{D} h_\mathrm{R}({\bf r})\int  D h_S^{i j}({\bf r}) \nonumber\\
&\times& \exp \left\{-\chi \int d^3r \, \rho_\mathrm{R}({\bf r})\rho_\mathrm{R}({\bf r})  
- \frac{v}{2}\int d^3r \left[\rho_\mathrm{C}({\bf r}) + \rho_\mathrm{R}({\bf r})\right]^2\right.
\nonumber\\
&-& \frac{w}{3!}\int d^3r \left[\rho_\mathrm{C}({\bf r}) + \rho_\mathrm{R}({\bf r})\right]^3- g \int d^3r 
\, {\rm Tr}\left[S^{i j}({\bf r})S^{i j}({\bf r})\right]\nonumber\\
&+& \left.i\int d^3 r \, \rho_\mathrm{C}({\bf r}) h_\mathrm{C}({\bf r}) + i\int d^3 r \,
\rho_\mathrm{R}({\bf r}) h_\mathrm{R}({\bf r}) + i\int d^3 r \, S^{i j}({\bf r}) h_S^{i j}({\bf r})\right\}
\nonumber\\
&\times&  Z^{(0)} (\{N_n\},
K;\left[h_\mathrm{C}\right],\left[h_\mathrm{R}\right],\left[h_S^{i j}\right])   \quad,
\label{Partition_func}
\end{eqnarray}
where the partition function $Z^{(0)}$ of the non - interacting system in the external fields $h_\mathrm{C}$, $h_\mathrm{R}$, $h^{ij}_S$.  is given by 
\begin{eqnarray}
&&Z^{(0)} (\{N_n\}, K;\left[h_\mathrm{C}\right],\left[h_\mathrm{R}\right],\left[h_S^{ij}\right])
\nonumber\\
&&\qquad\quad=\,\, \int \prod_{n=1}^{K} \mathcal{D}{\bf r}_n(s) \, d^3R_n \, d^2 u_n \, \delta
(|{\bf u}_n| - 1) \delta({\bf r}_n(f_nN_n) - {\bf R}_n)
\nonumber\\
&&\qquad\qquad\times\,\,\exp \left\{-
  \frac{3}{2b^2}\int_{0}^{f_n N_n} d s \left(\frac{\partial{\bf
        r}_n}{\partial s}\right)^2 - i \int_{0}^{f_n N_n} d s \, h_\mathrm{C} ({\bf r}_n(s))\right.
\nonumber\\
&&\qquad\qquad -\,\, i \int_{0}^{(1 - f_n)
  N_n} d s \, h_\mathrm{R}({\bf R}_n + {\bf u}_n s)
\nonumber\\
&&\qquad\qquad\left. -\,\, i \int_{0}^{(1 - f_n) N_n} d s \, h_S^{i j}({\bf R}_n + {\bf u}_n s) 
 \left(u_n^i u_n^j - \frac{1}{3}\delta^{i j}\right)\right\}.
\label{Partition_free}
\end{eqnarray}

The composition of the system is assumed to be equilibrated with respect to the  total number of stiff ($N_{\mathrm R}$) and flexible ($N_{\mathrm C}$) segments as well as to the number of junction points ($N_{\mathrm J}$) between stiff and flexible parts . Therefore the description can be reduced, i.e. $\{N_n\}, K \rightarrow N_{\mathrm R}, N_{\mathrm C}, N_{\mathrm J}$, and it is more convenient to switch to  a grand canonical partition function
$Z$ of the interacting polymer  written in terms of the grand canonical partition function 
$Z^{\mathrm (0)}$ of the non-interacting polymer in the external fields $h_\mathrm{C}$, $h_\mathrm{R}$, $h^{ij}_S$, i.e.
\begin{eqnarray}
Z(\mu,\epsilon,\sigma) &=& \int \mathcal{D}\rho_\mathrm{C}({\bf r}) \int \mathcal{D}\rho_\mathrm{R}({\bf r})\int
D S^{i j}({\bf r}) \int \mathcal{D} h_\mathrm{C}({\bf r})
\int \mathcal{D} h_\mathrm{R}({\bf r})\int  D h_S^{i j}({\bf r}) \nonumber\\
&\times& \exp\left\{-\beta H_{\rm int}\left[\rho_{\rm C}, \rho_{\rm R}\right]
+ i\int \mathrm{d}^3 r \, \rho_\mathrm{C}({\bf r}) h_\mathrm{C}({\bf r}) + i\int \mathrm{d}^3 r \,
\rho_\mathrm{R}({\bf r}) h_\mathrm{R}({\bf r}) + i\int \mathrm{d}^3 r \, S^{i j}({\bf r}) h_S^{i j}({\bf r})\right\}
\nonumber\\
&\times& Z^{(0)} (\mu,\epsilon,\sigma;\left[h_\mathrm{C}\right],\left[h_\mathrm{R}\right],\left[h_S^{i j}\right]).
\label{Partition1}
\end{eqnarray}
In Eq.(\ref{Partition1})  $\mu$ denotes the chemical potential conjugated to the whole number of segments and $-\epsilon$ is the energy gain of a helical segments 
compared to coil segment. The meaning of the cooperativity parameter $\sigma$ was discussed in the introduction.

The grand canonical partition function $Z^{(0)}$ of the non-interacting system can be derived by using the polymeric correlation function $\Xi^{(0)}$, i.e. $Z^{(0)} (\mu,\epsilon,\sigma;\left[h_\mathrm{C}\right],\left[h_\mathrm{R}\right],\left[h_S^{i j}\right]) = \int d 1 d 1' \: \Xi^{(0)} (1, 1';\mu,\epsilon,\sigma; \left[h_\mathrm{C}\right],\left[h_\mathrm{R}\right],\left[h_S^{i j}\right])$. The polymeric correlation function $\Xi^{(0)} (1, 1';\mu,\epsilon,\sigma; \left[h_\mathrm{C}\right],\left[h_\mathrm{R}\right],\left[h_S^{i j}\right])$ gives the conditional unnormalized probability of finding the first segment of the copolymer at the coordinate $1$ provided that the last segment is at $1'$. The symbol $1$ stands either for ${\bf r}_1$ or for $({\bf r}_1, {\bf u}_1)$ depending on whether the first segment is a flexible or a stiff one. The same holds for the coordinate $1'$ of the last segment. The polymeric correlation function is therefore the partition function of the multiblock helix-coil copolymer with two ends fixed at $1$ and $1'$.

By using the equations of motion for the rod and coil Green functions - $G_{\mathrm{coil}}$ and 
$G_{\mathrm{rod}}$ - the polymeric correlation  function $\Xi^{\mathrm (0)}$ can be constructed. 
The inverse Greens function operators are given by \cite{Nowak2} 
\begin{eqnarray}
\widehat{G}_{\rm coil}^{-1} &=& \delta({\bf r} - {\bf r}') \left( \beta\mu -
  \frac{b^2}{6} \nabla_r^2 + ih_\mathrm{C}({\bf r}) \right)\label{GG1}\\
\widehat{G}_{\rm rod}^{-1}  &=&  \delta({\bf r} - {\bf r}')\left(\beta(\mu - \epsilon) - \frac{b^2 ({\bf u}\cdot {\nabla_\mathrm{R}})^2}{\beta(\mu -
    \epsilon)} + ih_\mathrm{R}({\bf r}) + i\overset{\leftrightarrow}{h}_{S}({\bf r}) :
\overset{\leftrightarrow}{P} \right),
\label{GG2}
\end{eqnarray}
where the tensor $\overset{\leftrightarrow}{P}$ is given by $P^{i j} \equiv u^{i} u^{j} - \delta^{i j}/3$.
 In Eqs.(\ref{GG1}) - (\ref{GG2}) we have set the elementary unit lengths in coil and rod parts equal to each other (namely to $b$) for simplicity.  This affects some quantitative values but definitely does not alter qualitative predictions of this paper.

\begin{figure}[h!]
\begin{center}
\includegraphics[width=0.83\linewidth]{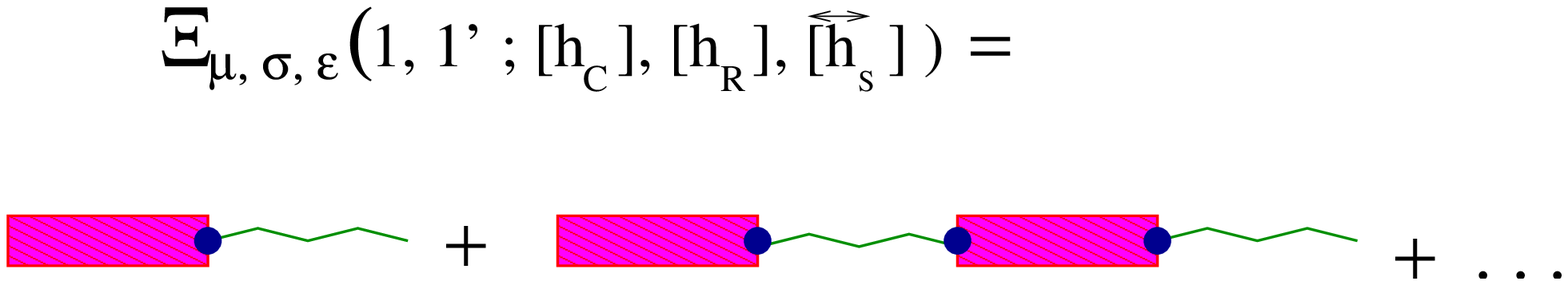}       
\caption[Geometric progression of rod and coil Green function operators]{Pictorial representation of the
        geometric progression in Eq.(\ref{Progres}). 
	One rod-coil unit constitutes the basic building block.}
\label{Geometry}
\end{center}
\end{figure}

Knowing the inverse Green function operators the grand canonical polymeric correlation
function can now be represented as the following Gaussian 2-dimensional path integral
\begin{eqnarray}
&&\Xi^{(0)}(1,1';\mu,\epsilon,\sigma;\left[h_\mathrm{C}\right],\left[h_\mathrm{R}\right],\left[h_{S}^{i j}\right])
=  \frac{1}{\Theta} \int \mathcal{D} \psi \mathcal{D} \varphi \: \psi(1) \: \varphi(1')
\nonumber\\
&&\qquad\qquad\quad\qquad  \times \exp \left\{ - \frac{1}{2} \int \mathrm{d} 3 \, \mathrm{d} 4 \left(
\begin{array}{c}
\psi(3)\\
\varphi(3)
\end{array}\right)^{\mathsf T}\left(
\begin{array}{cc}
\widehat{G}_{\rm rod}^{-1}& - \sigma^{1/2}\\
-\sigma^{1/2}& \widehat{G}_{\rm coil}^{-1} 
\end{array}\right)\left(
\begin{array}{c}
\psi(4)\\
\varphi(4)
\end{array}\right)\right\},
\nonumber\\
\label{Path_int}
\end{eqnarray}
where 
\begin{eqnarray}
\Theta = \int \mathcal{D} \psi \mathcal{D} \varphi \exp \left\{ - \frac{1}{2} \int \mathrm{d} 3 \,\mathrm{d} 4 \left(
\begin{array}{c}
\psi(3)\\
\varphi(3)
\end{array}\right)^{\mathsf T}\left(
\begin{array}{cc}
\widehat{G}_{\rm rod}^{-1}& - \sigma^{1/2}\\
-\sigma^{1/2}& \widehat{G}_{\rm coil}^{-1} 
\end{array}\right)\left(
\begin{array}{c}
\psi(4)\\
\varphi(4)
\end{array}\right)\right\}.
\nonumber\\
\end{eqnarray}
The choice of $\psi(1) \: \varphi(1')$ under the path integral yields one rod-coil unit
as a basic building block, see Fig.(\ref{Geometry}) and Eq.(\ref{Progres}). This choice assigns 
the coordinates $(1)$ and $(1')$ to $(\mathbf{r},\mathbf{u})$ and $(\mathbf{r'})$ respectively.
Similarly, $\psi(3)$ and $\psi(4)$  are shorthand notations for $\psi (\mathbf{r}_3,\mathbf{u}_3)$ and $\psi (\mathbf{r}_4,\mathbf{u}_4)$.

\begin{figure}[ht]
\begin{center}
\includegraphics[width=0.55\linewidth]{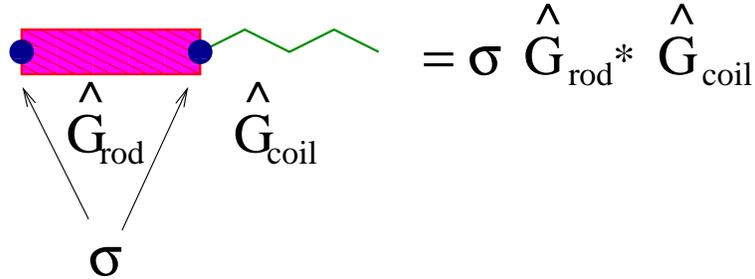}
\caption[One rod-coil building block]{Pictorial representation of the
       series ratio in Eq.(\ref{Progres}): each bar (rod), zigzag line (coil) and 
      fat dot correspond  to 
      $\widehat{G}_{\rm rod},\widehat{G}_{\rm coil} $ , and
      $\sigma^{1/2}$  respectively.}
\label{Ratio}
\end{center}
\end{figure}

The inversion of the $2 \times 2$-matrix in
Eq.(\ref{Path_int}) reads
\begin{eqnarray}
\left(
\begin{array}{cc}
\widehat{G}_{\rm rod}^{-1}& - \sigma^{1/2}\\
-\sigma^{1/2}& \widehat{G}_{\rm coil}^{-1} 
\end{array}\right)^{ - 1} = \frac{1}{\widehat{G}_{\rm rod}^{-1}\ast
\widehat{G}_{\rm coil}^{-1} - \sigma}
\left(
\begin{array}{cc}
\widehat{G}_{\rm coil}^{-1}&  \sigma^{1/2}\\
\sigma^{1/2}& \widehat{G}_{\rm rod}^{-1} 
\end{array}\right).
\end{eqnarray}
The calculation of the path integral in Eq.(\ref{Path_int}) yields the following result
\begin{eqnarray}
\Xi^{(0)}(1,1';\mu,\epsilon,\sigma;\left[h_\mathrm{C}\right],\left[h_\mathrm{R}\right],\left[h_{S}^{i
    j}\right]) &=& \frac{\sigma^{1/2}}{\widehat{G}_{\rm rod}^{-1}\ast
  \widehat{G}_{\rm coil}^{-1} - \sigma}\nonumber\\
&=& \sigma^{1/2}{\widehat{G}_{\rm rod} \ast\widehat{G}_{\rm
    coil}} \ast \Bigl[\widehat{1} + \sigma\widehat{G}_{\rm rod} \ast\widehat{G}_{\rm
    coil}\nonumber\\
 &+& \sigma^{2}\widehat{G}_{\rm rod} \ast\widehat{G}_{\rm
    coil}\ast \widehat{G}_{\rm rod}\ast\widehat{G}_{\rm coil}  + \dots \Bigr]. 
\nonumber\\
\label{Progres}
\end{eqnarray}
The asterisk in Eq.(\ref{Progres}) is a shorthand notation for a
convolution of Green function operators. The geometric progression
with a convolution as a binary relation has a clear pictorial
representation, see Fig.(\ref{Geometry}). 
The first term of this series has the analytical expression
$\sigma^{1/2}{\widehat{G}_{\rm rod} \ast\widehat{G}_{\rm coil}}$, 
whereas the ratio is equal to $\sigma{\widehat{G}_{\rm rod} \ast\widehat{G}_{\rm
    coil}}$, see also Fig.(\ref{Ratio}). The first term represents one rod-coil unit with one junction 
between rod and coil, hence the factor $\sigma^{1/2}$. If an additional building block is added, two more junctions are created, hence the factor $\sigma$ in the ratio. This series gives a
correct representation of the grand canonical polymeric correlation function.

The denominator in Eq.(\ref{Path_int}) can be avoided by introducing de
Gennes' $n \to 0$ trick \cite{Kholod}. Consider the two $n$-component vector
fields $\{\psi_{\alpha}, \varphi_{\alpha}\}$, where $\alpha = 1, 2, \dots n$.
Then Eq.(\ref{Path_int})  can be formally rewritten as
\begin{eqnarray}
&&\Xi^{(0)}(1,1';\mu,\epsilon,\sigma;\left[h_\mathrm{C}\right],\left[h_\mathrm{R}\right],\left[h_{S}^{i
    j}\right]) = \lim_{n \to 0} \prod_{\alpha = 1}^{n} \int \mathcal{D}
\psi_{\alpha} \mathcal{D} \varphi_{\alpha} \: \psi_{1}(1)
\:\varphi_{1}(1')\nonumber\\
&&\quad\quad\times \exp \Bigl\{ - \frac{1}{2} \int \mathrm{d} 3 \mathrm{d} 4\, \sum_{\alpha = 1}^{n} \left(
\begin{array}{c}
\psi_{\alpha}(3)\\
\varphi_{\alpha}(3)
\end{array}\right)^{\mathsf T}\left(
\begin{array}{cc}
\widehat{G}_{\rm rod}^{-1}& - \sigma^{1/2}\\
-\sigma^{1/2}& \widehat{G}_{\rm coil}^{-1} 
\end{array}\right)\left(
\begin{array}{c}
\psi_{\alpha}(4)\\
\varphi_{\alpha}(4)
\end{array}\right)\Bigr\}.
\nonumber\\
\label{Trick} 
\end{eqnarray}
Integrations over the external fields $h_\mathrm{C}$, $h_\mathrm{R}$, $h^{ij}_S$, over the densities $\rho_\mathrm{C}({\bf r}),
\rho_\mathrm{R}({\bf r})$, $S^{i j}({\bf r})$ and over all endpoints $\{1,1'\}$ yield the final field
theoretic representation of the grand canonical partition function
\begin{eqnarray}
Z (\mu, \epsilon, \sigma) &=& \lim_{n \to 0} \prod_{\alpha = 1}^{n} \int \mathcal{D}
\psi_{\alpha} \mathcal{D} \varphi_{\alpha} \: \left[\int \mathrm{d} ^3r \, \mathrm{d} ^2 u \: \psi_{1}({\bf r}, {\bf u})\right]
\:\left[\int \mathrm{d} ^3 r'  \: \varphi_{1}({\bf r}')\right]\nonumber\\&\times& \exp \left\{-\frac{1}{2}\sum_{\alpha = 1}^{n} \int \mathrm{d} ^3r \, \mathrm{d} ^2u
 \: \psi_{\alpha}({\bf r}, {\bf u})\Bigl[\beta(\mu - \epsilon) - \frac{b^2 \left({\bf u} \cdot
\nabla_{r}\right)^2}{\beta(\mu - \epsilon)}\Bigr]\psi_{\alpha}({\bf r}, {\bf u})\right.
\nonumber\\
&-& \frac{1}{2}\sum_{\alpha = 1}^{n} \int   \mathrm{d}^3r \varphi_{\alpha}({\bf
  r}) \Bigl[\beta\mu - \frac{b^2}{6} \nabla_{r}^2 \Bigr]\varphi_{\alpha}({\bf
  r})\nonumber\\
&-& \frac{\chi}{4} \int \mathrm{d}^3r \left[\sum_{\alpha = 1}^{n}\int \mathrm{d}^2u \:
  \psi_{\alpha}^2 ({\bf r}, {\bf u})\right]^2
\nonumber\\ 
&-& \frac{v}{8}\int \mathrm{d}^3r 
\left[\sum_{\alpha = 1}^{n} \int   \mathrm{d}^2u \:
  \psi_{\alpha}^2 ({\bf r}, {\bf u}) + \sum_{\alpha' = 1}^{n}
  \varphi_{\alpha'}^2({\bf r})\right]^2\nonumber\\
&-& \frac{w}{48}\int \mathrm{d}^3r 
\left[\sum_{\alpha = 1}^{n} \int   \mathrm{d}^2u \:
  \psi_{\alpha}^2 ({\bf r}, {\bf u}) + \sum_{\alpha' = 1}^{n}
  \varphi_{\alpha'}^2({\bf r})\right]^3\nonumber\\&+& \sigma^{1/2} \sum_{\alpha = 1}^{n} \int \mathrm{d}^3r  \, \mathrm{d}^2u \: \psi_{\alpha}({\bf
  r}, {\bf u})\varphi_{\alpha}({\bf r})\nonumber\\
&-& \left.\frac{g}{4}  \sum_{\alpha = 1}^{n} \sum_{\alpha' = 1}^{n}  \int \mathrm{d}^3r \,
\mathrm{d}^2u \, \mathrm{d}^2u' \: P_{2} ( {\bf u}\cdot  {\bf u}') \: \psi_{\alpha}^2 ({\bf
  r}, {\bf u}) \: \psi_{\alpha'}^2 ({\bf r}, {\bf u}') \right\},
\label{Field_theory}
\end{eqnarray}
where the second Legendre polynomial is given by
\begin{eqnarray}
 P_{2} (\theta) = \frac{1}{2} \left(3 \cos^2 \theta - 1 \right).
\end{eqnarray}
To evaluate the partition function $Z (\mu, \epsilon, \sigma)$ the self-consistent field approximation is used \cite{grosberg0}.  
In the self-consistent field approximation fluctuations are neglected and the functional integral over the fields in 
Eq.(\ref{Field_theory}) is integrated by steepest descent.
The saddle point solutions for $\varphi_{\alpha}({\bf r})$ and $\psi_{\alpha}({\bf r}, {\bf u})$ are chosen such that the effective grand potential 
keeps the full symmetry of the Hamiltonian in replica space, i.e. it is invariant under rotations in replica space. This is the case for (see ref. \cite{Kholod})
\begin{eqnarray}
\psi_{\alpha} ({\bf r}, {\bf u}) &=& n_{\alpha} \psi ({\bf r}, {\bf u})\nonumber\\
\varphi_{\alpha}({\bf r}) &=& n_{\alpha} \varphi ({\bf r}),
\label{Symmetry}
\end{eqnarray}
where ${\bf n}$ is a unit vector such that $\sum_{\alpha = 1}^{n} n_{\alpha}^2 = 1$. 

In order to make the problem more tractable we expand $\psi({\bf r}, {\bf u})$ in spherical harmonics, $\psi({\bf r}, {\bf u}) = \sum_{l, m} \psi_{lm}({\bf r}) Y_{lm}({\bf u})$, (see e.g. ref. \cite{gray}).
Since the solution for $\psi(\mathbf{r},\mathbf{u})$ must respect uniaxial and cylindrical symmetry, 
the expansion reduces to Legendre polynomials, i.e. $m\equiv0$. This expansion is truncated to the lowest nontrivial order \cite{Nowak4}
\begin{eqnarray} 
\psi({\bf r}, {\bf u}) \approx \left(\frac{1}{4 \pi}\right)^{1/2}
\psi_{0}({\bf r}) + \left(\frac{5}{4 \pi}\right)^{1/2} \psi_{2}({\bf r}) P_{2} ({\bf u} \cdot {\bf n}_z),
\label{exp-legendre}
\end{eqnarray}
where ${\bf n} $ is the main direction along which the rods in the
core of the globule are aligned, if the system forms an anisotropic globule with aligned rods. 
If they are not aligned and the system forms 
an amorphous  globule, only the first term in the expansion in Eq.(\ref{exp-legendre}) differs from zero.
On the other hand, the second term in Eq.(\ref{exp-legendre}) is responsible for the nematic LC-order. The main direction of alignment  ${\bf n} $ can be chosen arbitrarily without loss of generality, 
since a change in alignment direction only corresponds to a rotation of the complete globule in the
laboratory coordinate frame.  Thus we choose ${\bf n} $ directed  along  the  $z$-axis of the $(x,y,z)$ laboratory frame, i.e. 
${\bf n} = {\bf n}_z$.  This expansion allows us to perform the ${\bf u}$-integrations.

The resulting  effective saddle point grand potential $\Omega(\mu, \epsilon, \sigma)$ can now be calculated. It is given 
in terms of the saddle point fields $\psi_{0}({\bf r})$, $\psi_{2}({\bf r})$ and $\varphi({\bf r})$ by
\begin{eqnarray}
\beta \Omega(\mu, \epsilon, \sigma) &=& \frac{\beta(\mu -\epsilon)}{2}\int \mathrm{d}^3 r \left[\psi^2_{0}({\bf r}) + \psi^2_{2}({\bf r}) \right] \nonumber\\&-&
\frac{b^2}{210\,\beta\left(\mu  -\epsilon \right)} \int \mathrm{d}^3 r \Bigl\{35\,\psi_{0}({\bf r}) \nabla_r^2 \psi_{0}({\bf r})
\nonumber\\&+&
14\,{\sqrt{5}}\,\psi_{0}({\bf r})\Bigl[2\,\partial_z^2 - \partial_x^2 - \partial_y^2 \Bigr]\psi_{2}({\bf r})
\nonumber\\
&+&\psi_{2}({\bf r})\Bigl[25\,\partial_x^2 + 25\,\partial_y^2 + 55\,\partial_z^2\Bigr]\psi_{2}({\bf r})\Bigr\}
\nonumber\\&+&
\frac{1}{2}\int   \mathrm{d}^3r \: \varphi({\bf r})\Bigl[ \beta\mu -  \frac{b^2}{6} \nabla_{r}^2 \Bigr] \varphi({\bf r}) 
 \nonumber\\&+&
\frac{\chi}{4}\int \mathrm{d}^3 r \left[ \psi^2_{0}({\bf r}) + \psi^2_{2}({\bf r}) \right]^2
\nonumber\\&+& \frac{v}{8}\int \mathrm{d}^3 r \left[ \varphi^2({\bf r})  + \psi^2_{0}({\bf r}) + \psi^2_{2}({\bf r}) \right]^2 
\nonumber\\&+& \frac{w}{48}\int \mathrm{d}^3 r\, \left[ \varphi^2({\bf r})  + \psi^2_{0}({\bf r}) + \psi^2_{2}({\bf r}) \right]^3
\nonumber\\&-& 2\,{\sqrt{\pi \sigma }}\int \mathrm{d}^3 r\, \varphi({\bf r})\, \psi_{0}({\bf r})
\nonumber\\&+&
\frac{g}{245}\int \mathrm{d}^3 r \left\{\,\psi_{2}({\bf r})\,\left[ 7\,\psi_{0}({\bf r}) + \sqrt{5}\,\psi_{2}({\bf r})\right]\right\}^2. 
\label{F-exp}
\end{eqnarray}
The coil and rod densities as well as the orientation density are given by the following relations 
\begin{eqnarray}
\rho_{C} ({\bf r}) &=& \frac{1}{2} \varphi^2 ({\bf r})\nonumber\\
\rho_{R} ({\bf r}) &=& \frac{1}{2} \int d^2 u \psi^2 ({\bf r},{\bf u} )\simeq \frac{1}{2} \left[ \psi_{0}^2 ({\bf r}) + \psi_{2}^2 ({\bf r})\right] \nonumber\\
S({\bf r}) \equiv  S^{z z} ({\bf r}) &=& \frac{1}{3} \int d^2 u P_2 (\cos \theta)   \psi^2 ({\bf r},{\bf u} )\nonumber\\
&\simeq& \frac{2}{\sqrt{5}} \psi_{2}({\bf r})\left[ \psi_{0}({\bf r}) + \sqrt{5} \psi_{2}({\bf r})\right].
\label{Densities}
\end{eqnarray}

Functional minimization of Eq.(\ref{F-exp}) with respect to $\psi_{0}({\bf r})$, $\psi_{2}({\bf r})$ and $\varphi({\bf r})$ yields 
a set of three coupled partial differential equations, which are solved numerically with the finite element tool kit Gascoigne~\cite{gascoigne}.
The results are extensively discussed in the next section.
\section{Results}
In this section the numerical results for the full set of equations describing 
a rod-coil multiblock copolymer with a variable composition of stiff and flexible segments are presented. 
The segment length is set to $b=1$. In addition, all energies such as 
$\epsilon$, $\mu$ and also the saddle point free energy $F$ are given in units of $k_\mathrm{B} T$. In this section
this will not be indicated  in order to avoid complicated notation.  
Eq.(\ref{F-exp}) describes the copolymer in a grand-canonical representation. In the grand-canonical 
ensemble the total number of segments of the polymer $N$ is not fixed but its mean value is determined by equilibrium conditions. 
 
\begin{figure}[h!]
\begin{center}
\includegraphics[width=0.7\linewidth]{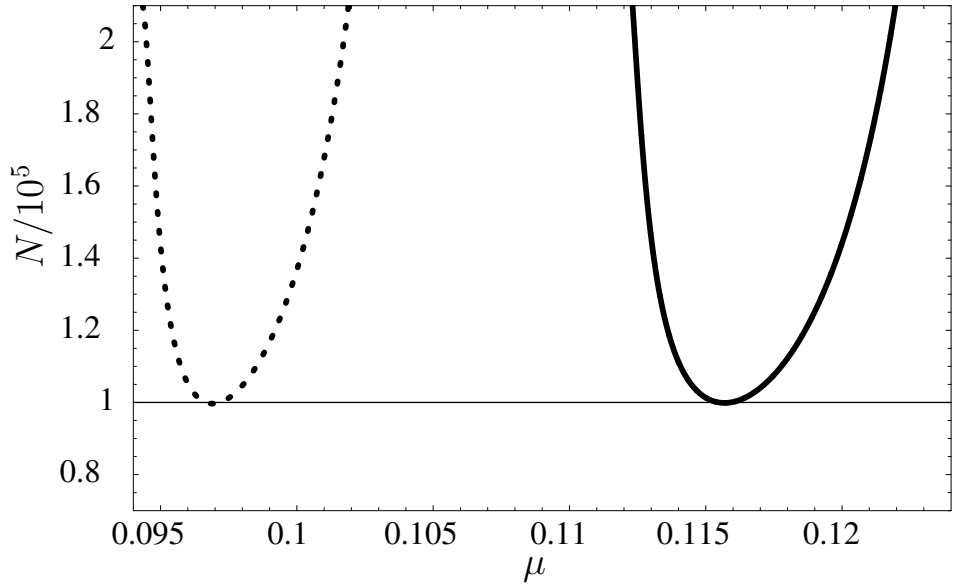}
\caption[Plot of $N$ versus $\mu$]{$N$ as a function of $\mu$ for $w=1$, $v = - 0.2$ and $\sigma=10^{-4}$. The dotted curve 
corresponds to $\epsilon=0.08$ and $\chi=0$. The continuous curve corresponds to $\epsilon=0.1$ and $\chi=0.0138$.}
\label{mu-N-rod-coil}
\end{center}
\end{figure}

A real polymer has a fixed length. In order to ensure this 
fixed length $N$, the chemical potential $\mu$ is - for each set of physical parameters 
($v,w,\chi,g,\epsilon,\sigma$) - tuned such that the equilibrium value of $N$ is equal to the desired one. 
The total number of segments $N = N_{\rm coil} + N_{\rm rod}$ is 
calculated by numerical integration over the rod and coil densities given by  Eq.(\ref{Densities}) . 
For a given set of parameters, $N(\mu)$ can be computed \cite{Nowak4}  and a typical example of this curve is shown in Fig.(\ref{mu-N-rod-coil}).
For $\mu \rightarrow 0$ the total number of segments $N$ diverges. This corresponds to the 
$N \sim \mu^{-1}$ behavior which is well-known for a $\Theta$-solvent chain \cite{Cloizeaux}. The divergence of $N$ at a specific 
value $\mu$ on the right hand side of the minimum corresponds to a fully collapsed infinite globule and has been discussed first by Kholodenko and Freed \cite{Kholod}. 
These two branches of $N(\mu)$ meet each other in the minimum of  $N(\mu)$ which can be associated with the coil-globule transition point, as 
will be shown in the next section.
Since $N$ is always fixed by tuning $\mu$, 
it is possible to distinguish from a plot like the one shown in Fig.(\ref{mu-N-rod-coil}) 
whether the system is left of the coil-globule transition 
point (i.e. in the open chain regime) or right of the transition point (i.e. in the globular 
regime). However, 
for fixed $N$, it is necessary to choose one of the two branches. Since this work focuses on the study 
of globular structures, the numerical calculations are always restricted to the right branch including the minimum. 
The self-consistent field theory is only expected to give good results for this branch 
since fluctuations are neglected.

The three-body interaction parameter $w$ 
is chosen to be $w=1$ throughout the entire paper and the two-body interaction constant $v$ is always negative to ensure 
that the system stays in the globular regime (up to the transition point). 

As a first step the pure homopolymer globule case \cite{Kholod} has been studied in order to test the numerical routines.  For this end one should simply set $\psi_0 = \psi_2 = 0$ in Eq.(\ref{F-exp}). Here we are not going to discuss these results in detail . It is pertinent only to note that the resulting numerical solution fully supports the following well known theoretical results. The critical value of the two-body interaction constant, $v$, scales as $|v_c| \sim N^{- 1/2}$, whereas the maximal globule density in the critical point behaves as $\rho_{\rm crit} \sim N^{- 1/2}$.

A short remark on the terminology that will be used below is necessary at this point.
The terms phase and transition will be used frequently although the system is a polymer of finite length. All transitions 
are therefore crossovers of finite width with continuous order parameters. 
It is nevertheless common now as applied to "soft matter" to use the 
the term "phase" to distinguish the different structural states of a polymer and to refer to the crossover between these states as 
a "transition". 
\subsection{Coil-globule transition}
\begin{figure}[h!]
\begin{center}
\includegraphics[width=0.7\linewidth]{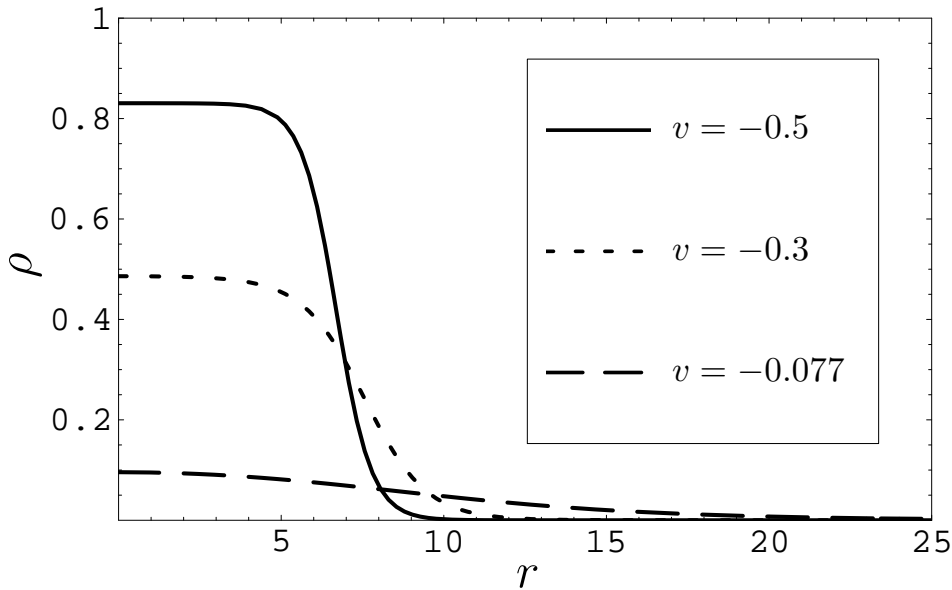}
\caption[Density profile of the copolymer]{This plot shows the radial density profile of the entire copolymer in radial direction for 
different values of $v$. The dashed curve for $v=-0.077$ corresponds to the coil-globule transition point.}
\label{dens-profile-coil-globule}
\end{center}
\end{figure}

The rod-coil multiblock copolymer shows a coil-globule transition similar to the one of a 
homopolymer. To demonstrate this, the interactions which are 
specific for the stiff segments are set to zero, that is $\chi = g = 0$. In addition, 
the energy gain per stiff segment is set to zero ($\epsilon = 0$) and it is assumed that there is no cooperativity
in the formation of stiff segments ($\sigma =1$). The two-body interaction constant $v$ is varied. The transition point between 
coil and globule is defined to be the minimum of the $N(\mu)$-curve as it was explained above. The length of the polymer is fixed at $N=550$. 
Fig.(\ref{dens-profile-coil-globule}) shows the profile of the total density $\rho({\bf r})$ of the copolymer as a function of radial distance from 
the center. The total density $\rho({\bf r})$ at each point is given by 
\begin{eqnarray}
\rho({\bf r})=\rho_{\rm C}({\bf r})+\rho_{\rm R}({\bf r})=
\frac{1}{2}\varphi^2({\bf r})+\frac{1}{2}\psi_0^2({\bf r})+\frac{1}{2}\psi_2^2({\bf r}).   
\end{eqnarray}
As can be seen from Fig.(\ref{dens-profile-coil-globule}), the density profile becomes broader with decreasing $|v|$. At 
$v=-0.5$ the copolymer is deep in the globular state with a big plateau of almost constant density and a rather small surface layer 
of decreasing density. At the transition point $v=-0.077$ the plateau basically vanished and the surface layer becomes very broad. 
\begin{figure}[h!]
\begin{center}
\includegraphics[width=0.95\linewidth]{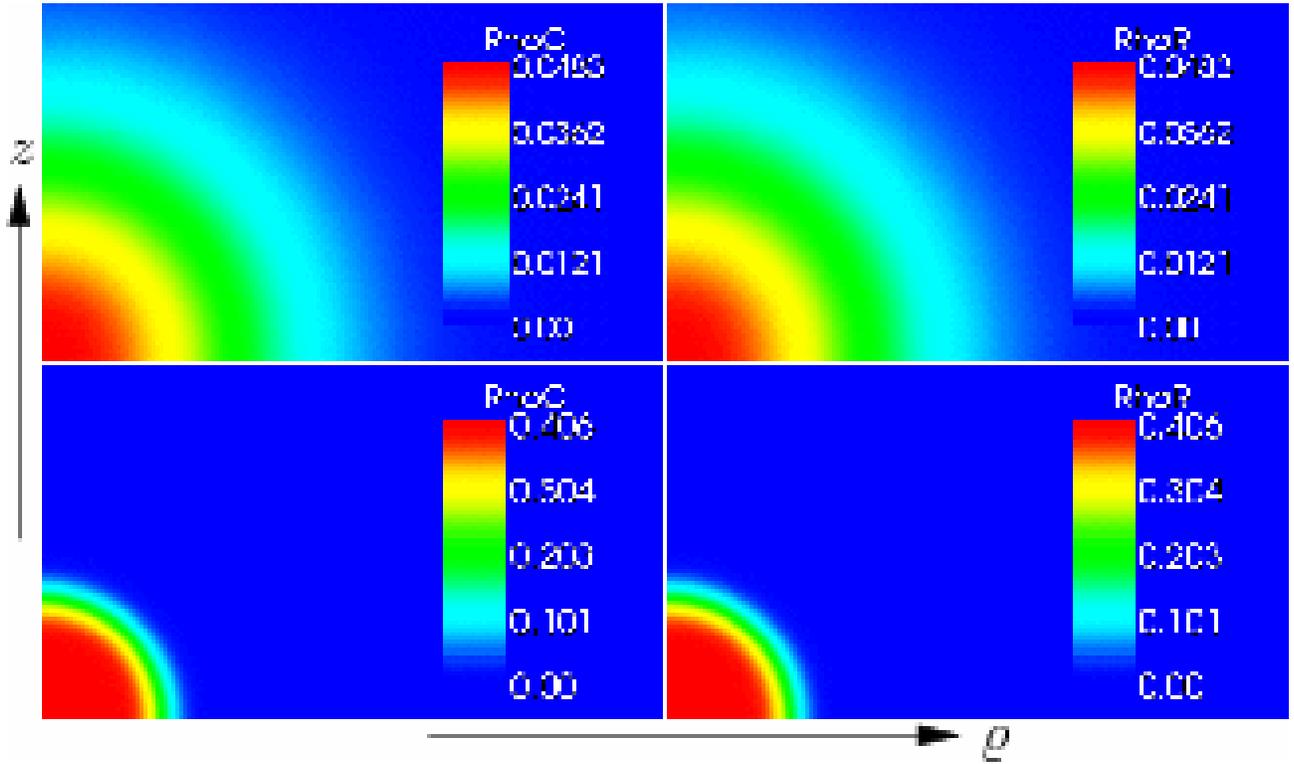}
\caption[Colour-coded plot of local density]{The density of flexible and stiff (helical) segments are  shown on the left and on the right panels respectively.  $v=-0.077$ corresponds  to the upper plots and $v=-0.5$ to the lower plots.}
\label{dens-colour-coil-globule}
\end{center}
\end{figure}
To further illustrate the structural change, Fig.(\ref{dens-colour-coil-globule}) shows a colour-coded plot of the local density 
in $\varrho$-$z$ space, where $\varrho$ denotes the radial direction and $z$ the axial direction in cylindrical coordinates. 
The center of the globule is located at the bottom left corner of each picture. The pictures on the left 
show the local density of flexible segments and the pictures on the right correspond to the local density of stiff segments. Red indicates high 
density and dark blue zero density. In the upper two plots the copolymer is at the transition point ($v=-0.077$). In the lower two plots 
it is deep in the globular state ($v=-0.5$).

\begin{figure}[h!]
\begin{center}
\psfrag{v}{\large $-v$}
\psfrag{R}{\large $R$}
\includegraphics[width=0.7\linewidth]{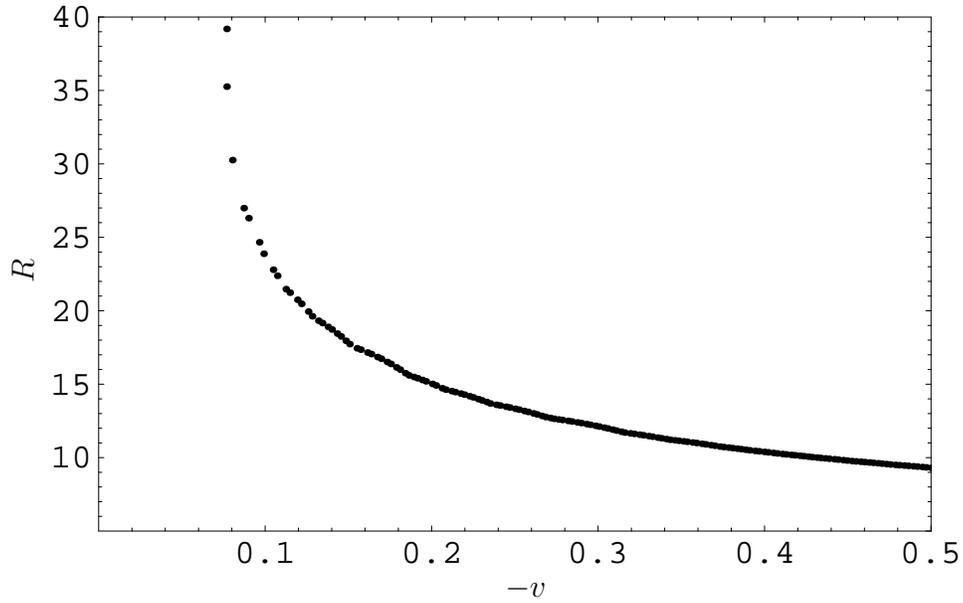}
\caption[Plot of globule radius $R$ versus $v$]{The globule radius $R$ is plotted as a function of $v$.}
\label{v-R-coil-globule}
\end{center}
\end{figure}

A much clearer indication that $v=-0.077$ corresponds indeed to a transition point can be seen 
from Fig.(\ref{v-R-coil-globule}), where the globule radius is plotted versus $v$. The globule radius is defined as the point $R$ in radial 
direction at which the density $\rho(R)$ has decreased to $\rho(R)=10^{-3}\rho_0$, where $\rho_0$ is the maximum density at the center of the globule. 
The radius $R$ shows a rapid increase when $v=-0.077$ 
is approached. Note, that the copolymer is finite (here: $N$=550) and therefore all transitions are crossovers as discussed above.
\subsection{Helix-coil transition}
In this section the fraction of stiff segments is investigated as a function of the 
energy gain per stiff segment $\epsilon$. This is similar to the  
helix-coil transition described by the Zimm-Bragg model \cite{zimm}. A major difference is that 
the model used here is a three-dimensional model of the polymer including interactions, 
 whereas the Zimm-Bragg model and its extensions are 
one-dimensional models with no three-dimensional interactions and no explicit entropy term . 
On the other hand, the Zimm-Bragg model can be solved exactly, whilst 
the self-consistent field treatment of the three-dimensional model is 
a mean-field approach which  neglects fluctuations.

Two different regimes will be discussed in the following: a low cooperativity 
regime with $\sigma$ in the range $0.05 - 1$ and a high cooperativity regime 
with $\sigma$ in the range $7\cdot10^{-3}-10^{-4}$.  
Remember, that $\sigma=1$ means no cooperativity and $\sigma=0$ means total cooperativity. 
Throughout this section $\chi$ and $g$ are set to zero. There are therefore no specific interactions between 
the stiff segments. The only interactions are attractive two-body interactions and repulsive 
three-body interactions between all segments. 
\begin{figure}[h!]
\begin{center}
\psfrag{h}{\large $\Theta_{\rm R}$}
\psfrag{eps}{\large $\epsilon$}
\psfrag{s1}{\small $\sigma=1.0$}
\psfrag{s07}{\small $\sigma=0.7$}
\psfrag{s05}{\small $\sigma=0.5$}
\psfrag{s02}{\small $\sigma=0.2$}
\psfrag{s005}{\small $\sigma=0.05$}
\includegraphics[width=0.9\linewidth]{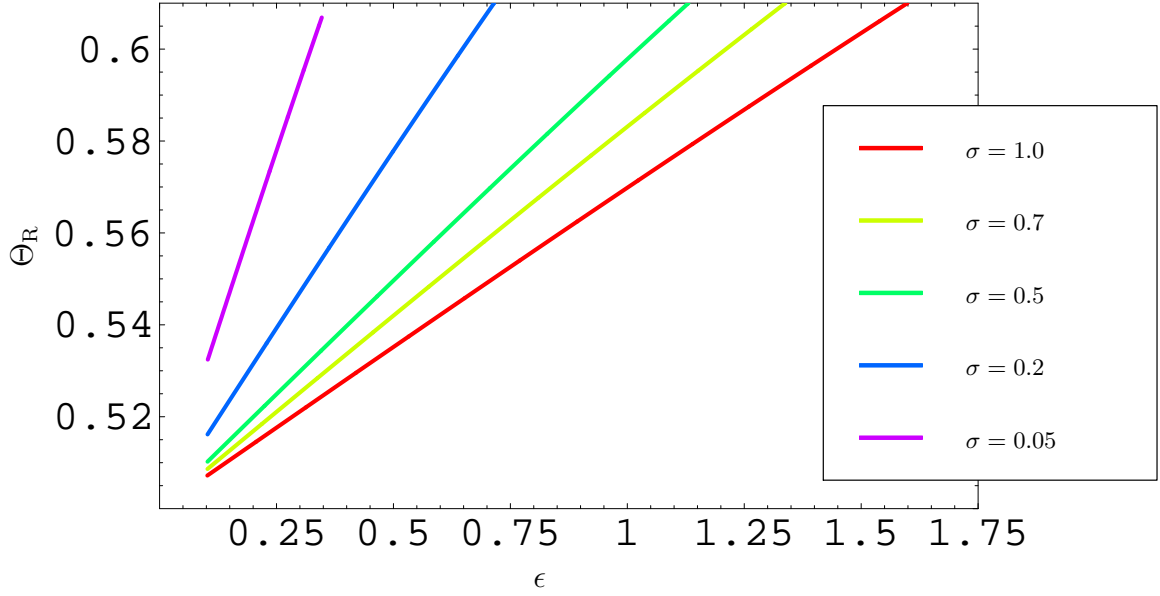}
\caption[Fraction of stiff segments versus $\epsilon$ for large $\sigma$]{The fraction of stiff segments $\Theta_{\rm R}$ is 
plotted as a function of energy gain $\epsilon$ per stiff segments for small (and zero) cooperativity.
$v$ = $-0.025$ and $N=2.5\cdot 10^4$.}
\label{eps-var-sighigh}
\end{center}
\end{figure}
Fig.(\ref{eps-var-sighigh}) shows how the fraction of stiff segments $\Theta_{\rm R}$ depends on $\epsilon$ for different values 
of $\sigma$ in the range $0.05 - 1$. Even for small cooperativity, the slope clearly depends on $\sigma$ and gets larger with 
increasing cooperativity (decreasing $\sigma$). 

\begin{figure}[h!]
\begin{center}
\psfrag{h}{\large $\Theta_{\rm R}$}
\psfrag{eps}{\large $\epsilon$}
\psfrag{sig1e4}{\small $\sigma=10^{-4}$}
\psfrag{sig5e4}{\small $\sigma=5\cdot10^{-4}$}
\psfrag{sig1e3}{\small $\sigma=10^{-3}$}
\psfrag{sig7e3}{\small $\sigma=7\cdot10^{-3}$}
\includegraphics[width=0.98\linewidth]{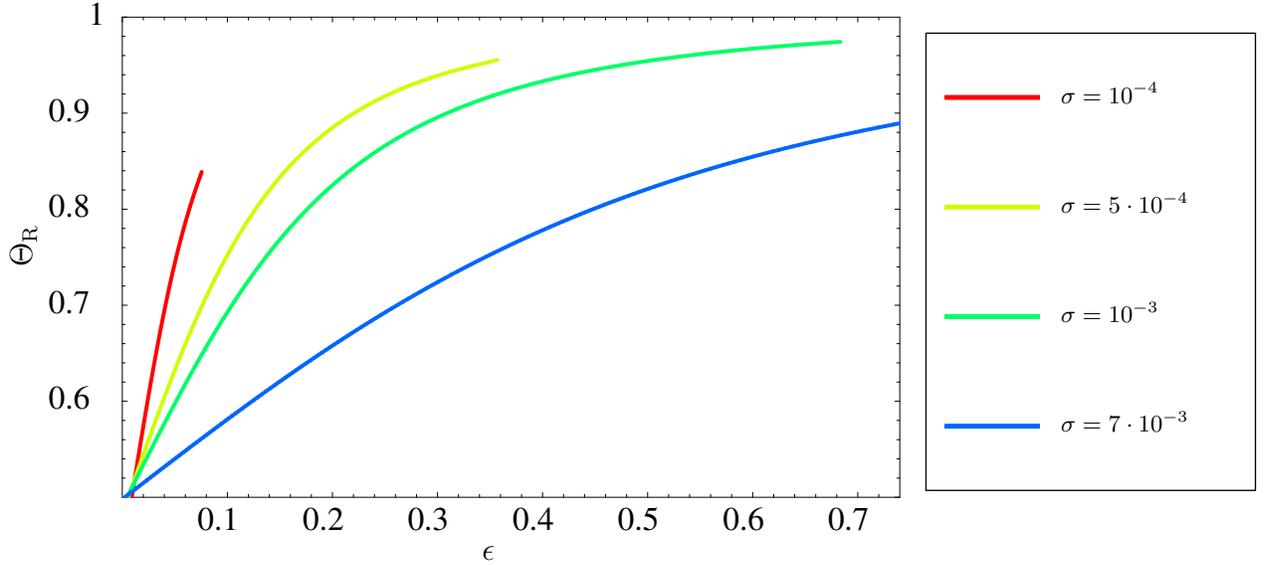}
\caption[Fraction of stiff segments versus $\epsilon$ for small $\sigma$]{The fraction of stiff segments $\Theta_{\rm R}$ is 
plotted as a function of $\epsilon$ for high cooperativity. $v$ = $-0.2$ and $N=10^5$.}
\label{eps-var-siglow}
\end{center}
\end{figure}

The high cooperativity regime is shown in Fig.(\ref{eps-var-siglow}).  
The slope of the $\Theta_{\rm R}(\epsilon)$-curves increases with decreasing $\sigma$ as expected. Especially for $\sigma=10^{-3}$ 
(green curve) it can be seen that the curve is asymptotically approaching $\Theta_{\rm R}=1$. The curves for $\sigma=10^{-4}$, $5\cdot10^{-4}$ 
and $10^{-3}$ 
end at a certain value of $\epsilon$. These values correspond to the coil-globule transition point. This can be explained as follows. 
If there is no additional selective interaction 
energy which favors a compactification of the stiff segments (i.e. $\chi=g=0$), the stiffening of parts of the chain due to an increase of 
$\Theta_{\rm R}$ with increasing $\epsilon$ pushes the chain segments further apart from each other and therefore leads to a more 
open structure. From a certain value of $\epsilon$ on, the system is thus pushed into the open chain regime. For higher cooperativity 
this effect is stronger, since the system forms less junctions points and hence on average longer rods. For smaller $\sigma$ the 
 coil-globule transition point is therefore reached at smaller values of $\Theta_{\rm R}$ and $\epsilon$. 

\begin{figure}[h!]
\begin{center}
\psfrag{l}{\large $L_{\rm R}$}
\psfrag{e}{\large $\epsilon$}
\includegraphics[width=0.65\linewidth]{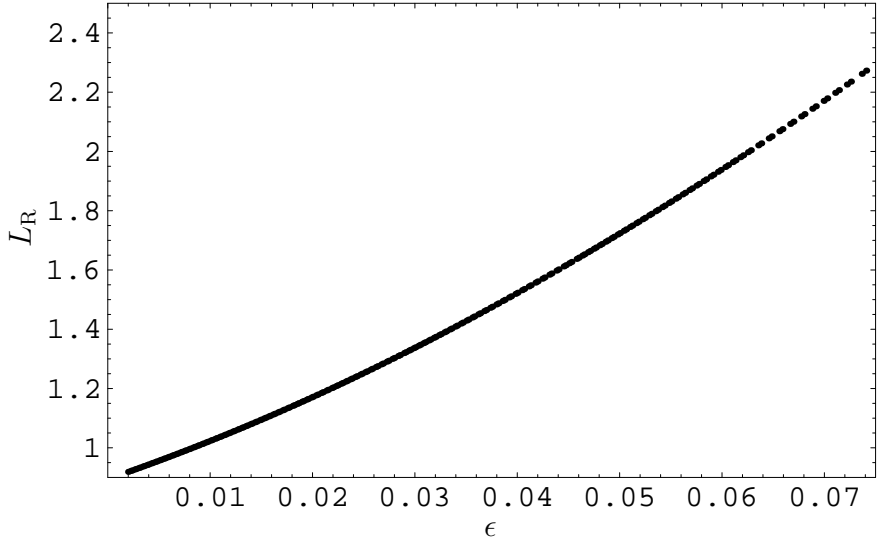}
\caption[Average rod length versus $\epsilon$]{The average rod length $L_{\rm R}$ is plotted as a function of $\epsilon$ for $\sigma=10^{-4}$, 
$v$ = $-0.2$ and $N=10^5$.}
\label{LR-eps}
\end{center}
\end{figure}

To illustrate this stiffening, in Fig.(\ref{LR-eps}) the average rod length $L_{\rm R}$ is plotted as a function of $\epsilon$ for 
$\sigma=10^{-4}$. 
In units of $b$ the average rod length $L_{\rm R}$ is given by the total number of stiff segments divided by half the total number  
of junction points between stiff rod and flexible chain. In the self-consistent field theory approach 
$L_{\rm R}$ is given by
\begin{eqnarray}
\frac{L_{\rm R}}{b} = \frac{\int_{0}^{\infty} d\varrho\,\varrho \int_{-\infty}^{\infty} dz\, \left[\psi^2_{0}(\varrho,z) 
+ \psi^2_{2}(\varrho,z) \right]}
{\int_{0}^{\infty} d\varrho\,\varrho \int_{-\infty}^{\infty} dz\,\, \varphi(\varrho,z)\, \psi_{0}(\varrho,z)}.
\end{eqnarray}
Fig.(\ref{LR-eps}) shows that for $\epsilon=0$ the average rod length is roughly equal to $1$. For $\epsilon>0$ the cooperativity effect 
sets in and the average rod length increases up to  $L_{\rm R}\approx 2.3$. At this value the stiffening of parts of the chain 
is strong enough to drive the polymer in the open chain regime. 

\begin{figure}[h!]
\begin{center}
\psfrag{h}{\large $\Theta_{\rm R}$}
\psfrag{e}{\large $\epsilon$}
\includegraphics[width=0.7\linewidth]{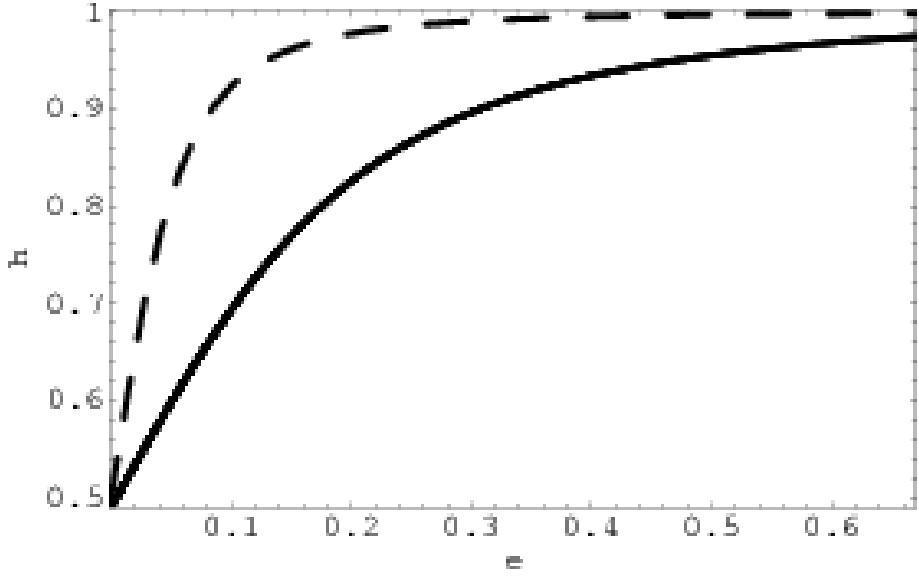}
\caption[$\Theta_{\rm R}$ versus $\epsilon$: comparison to Zimm-Bragg theory]{$\Theta_{\rm R}$ is plotted as a function of $\epsilon$ 
for $\sigma=10^{-3}$. The continuous curve shows the numerical result for $v=-0.2$ and $N=10^5$. The dashed curve shows the result of 
the one-dimensional Zimm-Bragg theory.}
\label{comp-BZ}
\end{center}
\end{figure}

It is of interest to compare the results of this three-dimensional model  with the 
exact one-dimensional Zimm-Bragg model discussed in Introduction. This is done in Fig.(\ref{comp-BZ}) for 
$\sigma=10^{-3}$. Note, that in the Zimm-Bragg language the $\epsilon$ used here corresponds to $-\Delta f = \ln s$. 
In the Zimm-Bragg model the increase of $\Theta_{\rm R}$ is steeper. 
This can be explained by the fact that in the three-dimensional model the entropy of the flexible and the stiff segments is explicitly taken into account, which hinders the generation of stiff segments. On the other hand, the mean-field character of the 
three-dimensional model might also weaken the cooperativity effect.
 
\subsection{Transition from amorphous to liquid-crystalline globule}
\begin{figure}[h!]
\begin{center}
\psfrag{h}{$\Theta_{\rm R}$}
\psfrag{c}{$-\chi$}
\psfrag{m}{$S$}
\includegraphics[width=0.49\linewidth]{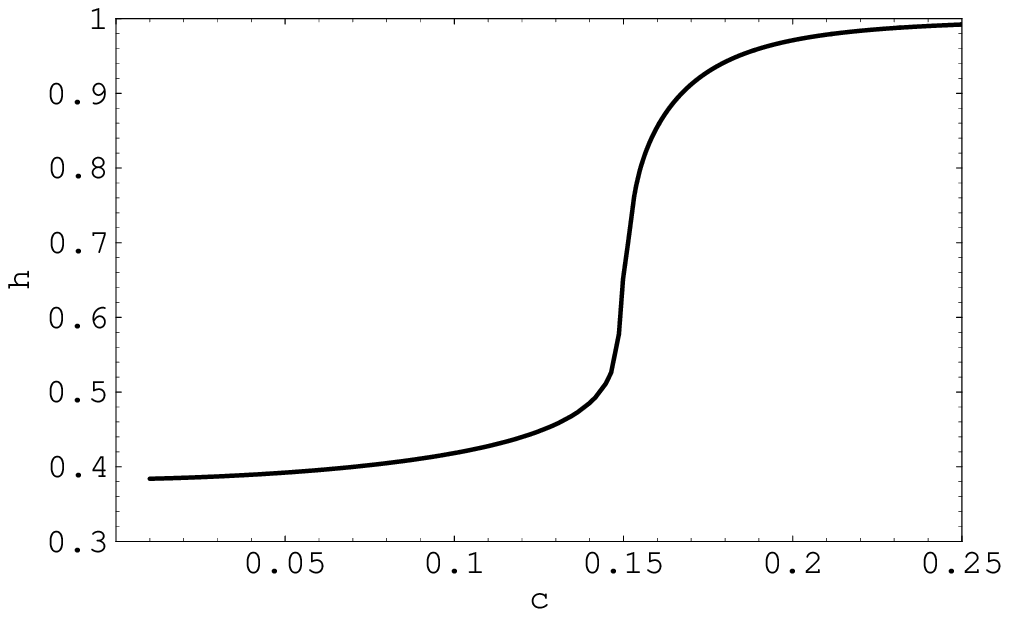}
\includegraphics[width=0.49\linewidth]{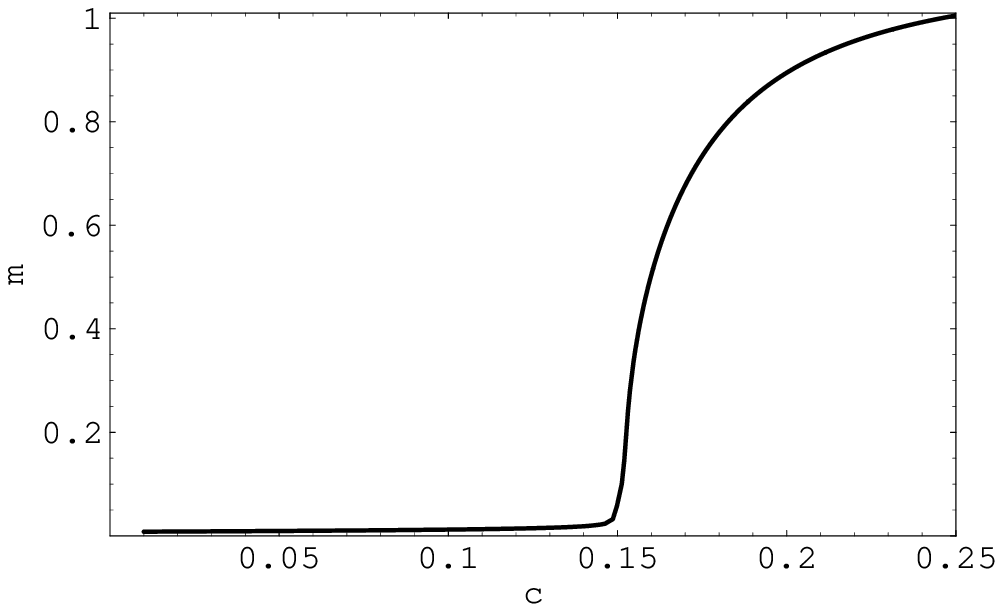}
\caption[Isotropic-nematic transition: $\Theta_{\rm R}$ and $S$ versus $\chi$]
{The fraction of stiff segments $\Theta_{\rm R}$ (left) and the nematic 
order parameter $S$ (right) are plotted as functions of $\chi$ for $\sigma=10^{-4}$, $N=9.5\cdot10^{3}$ and $\epsilon = g = 0$.}
\label{iso-nem1}
\end{center}
\end{figure}

The main focus of this work lies on the crossover from a disordered amorphous globule with a low or moderate fraction of stiff segments 
to an ordered liquid-crystalline globule with a very high fraction of stiff segments. In Fig.(\ref{iso-nem1}), on the left, the 
fraction of stiff segments $\Theta_{\rm R}$ is plotted as a function of the selective two-body interaction 
parameter $\chi$, which controls the strength of the additional interaction between stiff segments and therefore 
models selective solvent conditions (hydrophobicity). On the right the nematic order parameter $S$  
is plotted as a function of $\chi$. The nematic order parameter is given by 
\begin{eqnarray}
S &=& \frac{1}{3 N}\int \mathrm{d}^3 r \int \mathrm{d}^2 u \: P_2 
(\cos \theta) \psi^2({\bf r}, {\bf u})
\nonumber\\  
&=& \frac{2}{\sqrt{5} N}\int \mathrm{d}^3 r \:  \psi_{2}({\bf r})
\left[\psi_{0}({\bf r}) + \sqrt{5} \psi_{2}({\bf r})\right]. 
\end{eqnarray}
Throughout the entire section, the non-selective two-body interaction parameter is set to $v=-0.2$.
The two plots in Fig.(\ref{iso-nem1}) demonstrate that the onset of nematic 
order and the increase in the fraction of stiff segments occur simultaneously.

This transition occurs without an explicit angle-dependent alignment interaction, that is $g=0$.  
The transition is triggered by a subtle interplay of the entropy contribution (surface energy), represented by the derivative terms in Eq.(\ref{F-exp}), and bulk interaction energy, represented by the 
$\chi$-term. This surface energy has an entropic nature since the conformational set of surface segments is 
constrained. In a simple homopolymer globule it is isotropic~\cite{grosberg0}. 
For the rod-coil copolymer the surface energy is anisotropic and the surface tension in $\varrho$-direction 
is smaller than the one in $z$-direction. That is why the system tries to maximise its lateral surface in $\varrho$-direction and  
minimize it in $z$-direction, i.e. a nematic, cigar shaped, liquid-crystalline globule is formed.  

\begin{figure}[h!]
\begin{center}
\psfrag{z}{\Large $z$}
\psfrag{r}{\Large $\varrho$}
\includegraphics[width=0.9\linewidth]{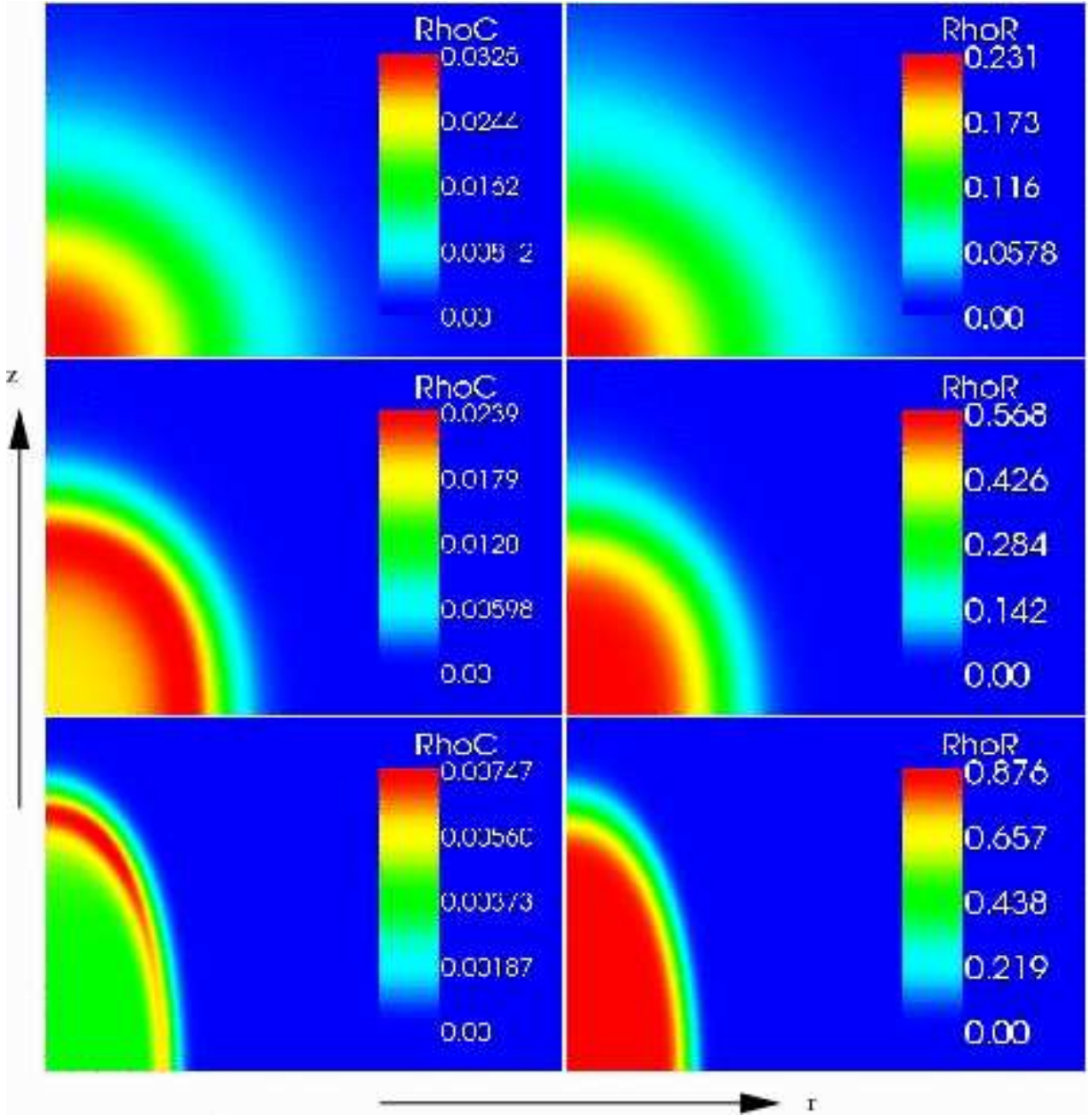}
\caption[Density plots for different $\chi$]{The density of the flexible segments is shown on the left and the density of the stiff
segments on the right. $\sigma=10^{-4}$, $\epsilon=0.1$ and $N=10^5$ for all plots. $\chi=-0.0138$ in the top line, $\chi=-0.0812$ 
in the middle line and $\chi=-0.18$ in the bottom line.}
\label{dens}
\end{center}
\end{figure}

To demonstrate how the shape of the globule changes during the transition, Fig.(\ref{dens}) shows  
a colour-coded plot of the local density in $\varrho$-$z$ space. 
The center of the globule is in the bottom left corner of each picture. $\varrho$ is 
increasing from left to right and $z$ is increasing from bottom to top.
Red indicates high density and dark blue zero density. In Fig.(\ref{dens}) a different set of parameters is chosen:
$\sigma=10^{-4}$, $N=10^5$ and $\epsilon=0.1$. Below, it will be discussed in detail how the nature of the crossover 
changes with $\sigma$, $\epsilon$ and $N$.  
The values of $\chi$ in Fig.(\ref{dens}) are chosen such that the top two pictures ($\chi=-0.0138$) show the system at the transition 
point between coil and globule, the middle two pictures ($\chi=-0.0812$) show the system at the transition point 
between amorphous and liquid-crystalline globule (as defined below) 
and the bottom two pictures ($\chi=-0.18$) show the system deep in the liquid-crystalline  
globule phase. The transition from amorphous to nematic liquid-crystalline globule is a crossover of finite width and it is a rather obvious choice, to define the transition point as the inflection point of the $S(\chi)$-curve in Fig.(\ref{iso-nem1}). 
At the transition point between coil and globule the system is spherical and has a very broad surface layer of decaying density. 
Although the density of the helical segments shown on the right is higher than the density 
of the flexible segments shown on the left, their distribution and the shape of the profile is very similar.  

At the transition point between amorphous and liquid-crystalline globule
the system adopts a slightly cylindrical shape indicating the onset of nematic order. It can also be seen that the 
density maximum of the flexible segments is not in the center of the globule denoting a repulsion of flexible 
segments from the center to the surface layer. The surface layer is now much narrower. 
Deep in the liquid-crystalline  
phase the globule has developed a strongly asymmetric cylindrical shape indicating strong 
nematic order. The repulsion of flexible segments from the center towards the surface layer can be 
seen clearly and the surface layer is now very narrow.
\subsubsection{$N$-dependence}
\begin{figure}[h!]
\begin{center}
\psfrag{h}{\large $\Theta_{\rm R}$}
\psfrag{c}{\large $-\chi$}
\psfrag{N1e4}{\small $N=9.5\cdot10^3$}
\psfrag{N3e4}{\small $N=3\cdot10^4$}
\psfrag{N1e5}{\small $N=10^5$}
\psfrag{N5e5}{\small $N=5\cdot10^5$}
\includegraphics[width=0.85\linewidth]{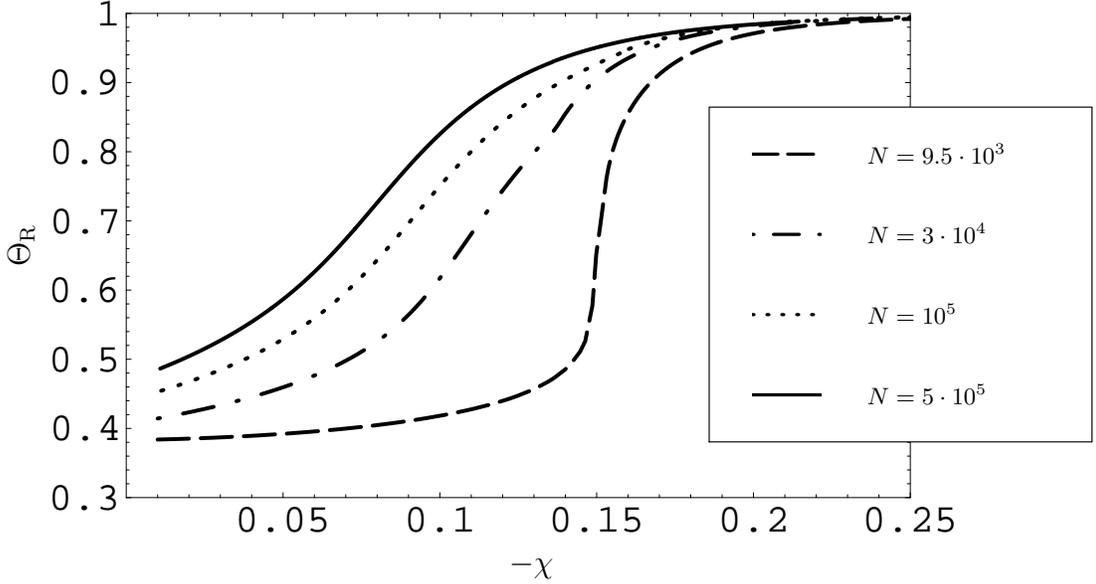}
\caption[$\Theta_{\rm R}$ versus $\chi$ for different $N$]{The fraction of stiff segments $\Theta_{\rm R}$ is plotted as a function 
of $\chi$ for different chain lengths. The crossover becomes sharper with decreasing chain length.}
\label{fracH-N}
\end{center}
\end{figure}

In this subsection we study  how the total chain 
length $N$ influences the transition from amorphous to liquid-crystalline globule.
Fig.(\ref{fracH-N}) shows the fraction of stiff segments as a function of $\chi$ for four different 
chain lengths. The crossover from an amorphous globule with moderate number of stiff segments 
to a liquid-crystalline globule with very high number of a stiff segments becomes sharper with decreasing chain length, 
which, at first sight, is a rather unusual and surprising behavior. 

\begin{figure}[h!]
\begin{center}
\psfrag{m}{\large $S$}
\psfrag{c}{\large $-\chi$}
\psfrag{N1e4}{\small $N=9.5\cdot10^3$}
\psfrag{N3e4}{\small $N=3\cdot10^4$}
\psfrag{N1e5}{\small $N=10^5$}
\psfrag{N5e5}{\small $N=5\cdot10^5$}
\includegraphics[width=0.77\linewidth]{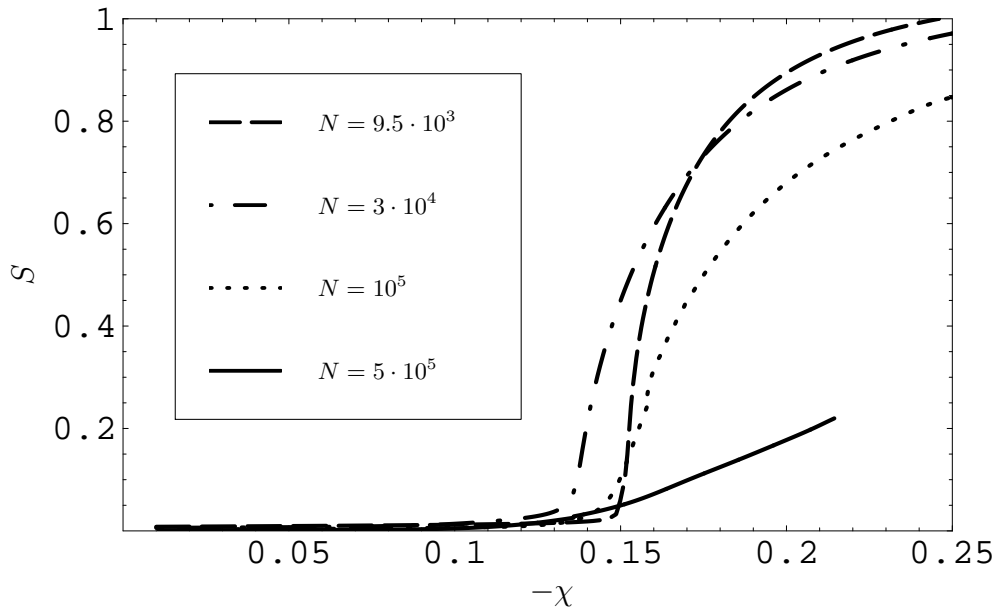}
\caption[$S$ versus $\chi$ for different $N$]{The nematic order parameter $S$ is plotted as a function 
of $\chi$ for different chain lengths.}
\label{S-N}
\end{center}
\end{figure}

Fig.(\ref{S-N}) shows the nematic order parameter $S$ as a function of $\chi$ for the four chain lengths. It can be seen 
that the increase in the order parameter is stronger for shorter chains. For $N=5\cdot10^5$ and $\chi=-0.214$ the system is already 
so deep in the globular phase that the corresponding value of $\mu$, which keeps the chain length fixed 
at $N=5\cdot10^5$, is very close to 
the value at which $N(\mu)$ diverges (see Fig.(\ref{mu-N-rod-coil})).  
The tuning of the chemical potential $\mu$ to ensure fixed chain length $N$ becomes therefore  
numerically impossible for higher values of $|\chi|$. Although it might be difficult to see, 
the corresponding $\Theta_{\rm R}(\chi)$-curve in Fig.(\ref{fracH-N}) also ends at $\chi=-0.214$. 

\begin{figure}[h!]
\begin{center}
\psfrag{z}{\Large $z$}
\psfrag{r}{\Large $\varrho$}
\includegraphics[width=0.95\linewidth]{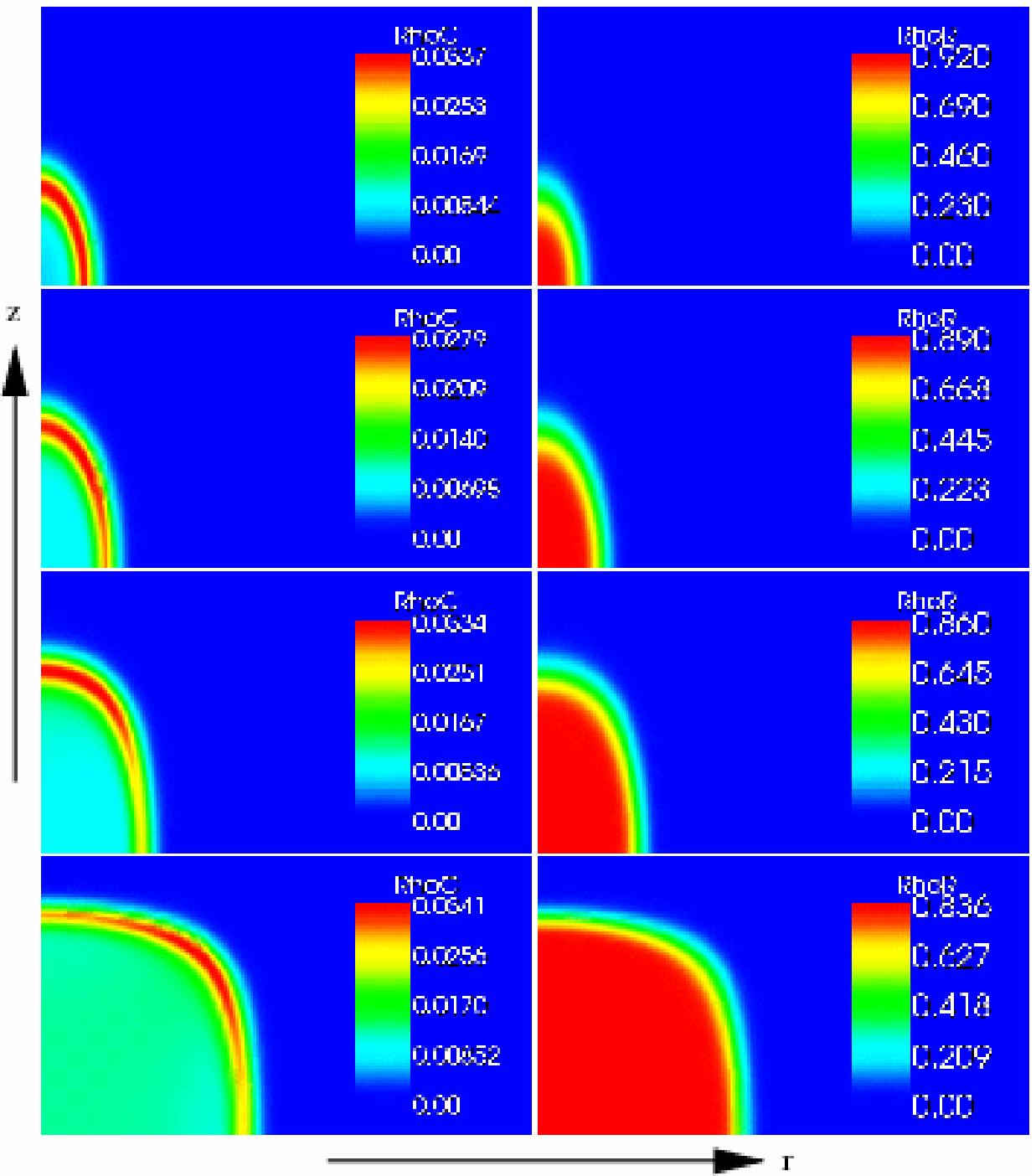}
\caption[Density plots for different $N$]{The density of the flexible segments is shown on the left and the density of the stiff
segments on the right. $\sigma=10^{-4}$, $\epsilon=g=0.0$ and $\chi=-0.18$ for all plots. From top to bottom $N$ increases as $10^4$, $3\cdot10^4$, 
$10^5$ and $5\cdot10^5$.}
\label{LC-N}
\end{center}
\end{figure}

To demonstrate how the shape of the liquid-crystalline globule changes with chain length $N$, Fig.(\ref{LC-N}) shows colour-coded
density plots of $\rho_{\rm C}$ and $\rho_{\rm R}$ for the four different values of the chain length at $\chi=-0.18$.  
The plots illustrate that the system changes from a cigar-like shape for $N=10^4$ towards an almost cylindrical shape for $N=5\cdot10^5$. 
Fig.(\ref{LC-N}) also demonstrates that the surface layer becomes smaller for larger systems, 
which indicates that a  larger system is deeper in the globular state at the same value of $\chi$ and $v$ but at the same time showing less 
nematic order. 

With increasing system size $N$ the liquid-crystalline globule phase eventually dies out (for $g=0$) because in the limit $N\rightarrow\infty$ the contribution of 
the surface terms in Eq.(\ref{F-exp}) vanishes. Since for $g=0$ these terms are the only non-spherical symmetric terms, the system 
cannot adopt a liquid-crystalline globular state any more. Without explicit alignment interaction a liquid-crystalline globule 
with nematic order can therefore only form in finite systems. The crossover becomes sharper for smaller systems because the 
entropic surface terms become more important compared to the isotropic interaction terms. The total value of the isotropic 
bulk interaction energy roughly scales as the volume of the globule, whilst the surface energy scales as the surface area of the globule.   
This shows that the transition to a liquid-crystalline polymer globule is actually driven by entropy, 
due to the entropic origin of the surface energy. 
 
\begin{figure}[h!]
\begin{center}
\psfrag{c}{\large $-\chi$}
\psfrag{L}{\large $L_{\rm R}$}
\psfrag{N4}{\small $N=10^4$}
\psfrag{N55}{\small $N=5\cdot10^5$}
\includegraphics[width=0.7\linewidth]{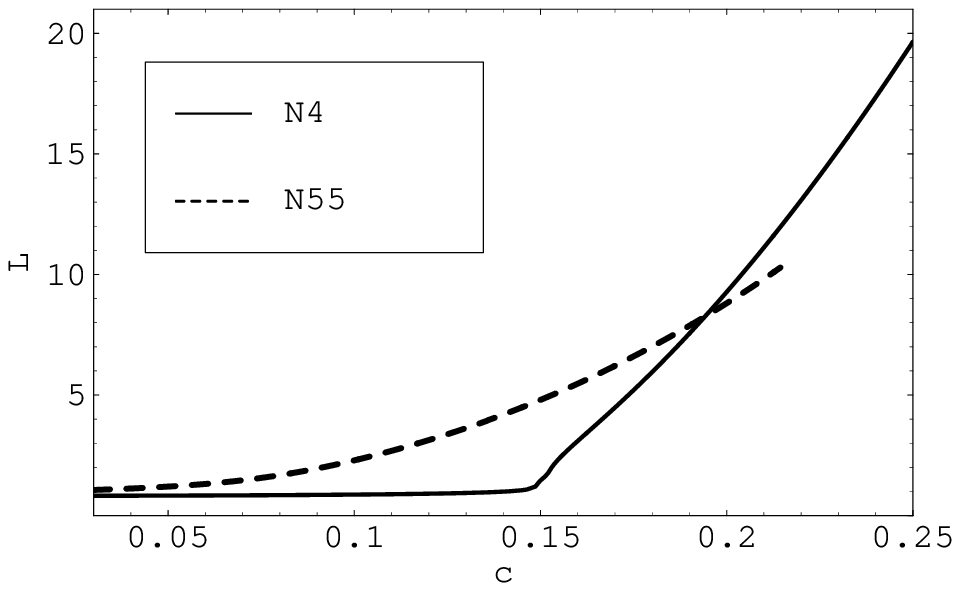}
\caption[$L_{\rm R}$ versus $\chi$ for different $N$]{The average length $L_{\rm R}$ of the stiff parts is plotted as 
a function of $\chi$ for $N=10^4$ and $N=5\cdot10^5$.}
\label{Lh-N}
\end{center}
\end{figure}

Finally, in Fig.(\ref{Lh-N}), the average length $L_{\rm R}$ of the stiff parts is plotted as a function of $\chi$ 
for $N=10^4$ and $N=5\cdot10^5$. The figure shows that the average length of the rods is equal to $1$ for very small 
values of $\chi$ and $\epsilon=0$. 
For $N=10^4$ it stays almost equal to $1$ until the transition point is reached and  
then shows a rather strong increase. For very high values of $\chi$ it should reach an asymptotic value. However, for $\chi>0.25$ the fraction of flexible segments ($1-\Theta_{\rm R}$, see Fig.(\ref{iso-nem1})) becomes unphysically small indicating that  the self-consistent field treatment reaches the limit of its validity. For $N=5\cdot10^5$ the crossover to a stronger increase in 
average rod length with $-\chi$ is much smoother analogous to $\Theta_{\rm R}$ and $S$ (see Figs.(\ref{fracH-N}, \ref{S-N})).  

The stronger increase in average rod length for shorter polymers corresponding to a sharper isotropic-nematic transition indicates that this 
transition is also a cooperative process. To study 
the influence of cooperativity, the dependence of the isotropic-nematic transition on $\sigma$ is discussed in the next subsection.
 
\subsubsection{$\sigma$-dependence}
The crossover from amorphous globule to liquid-crystalline globule is a cooperative process. 
\begin{figure}[h!]
\begin{center}
\psfrag{h}{\large $\Theta_{\rm R}$}
\psfrag{c}{\large $-\chi$}
\psfrag{sig1e3}{\scriptsize $\sigma=10^{-3}$}
\psfrag{sig5e4}{\scriptsize $\sigma=5\cdot10^{-4}$}
\psfrag{sig1e4}{\scriptsize $\sigma=10^{-4}$}
\psfrag{sig5e5}{\scriptsize $\sigma=5\cdot10^{-5}$}
\psfrag{sig1e5}{\scriptsize $\sigma=10^{-5}$}
\includegraphics[width=0.75\linewidth]{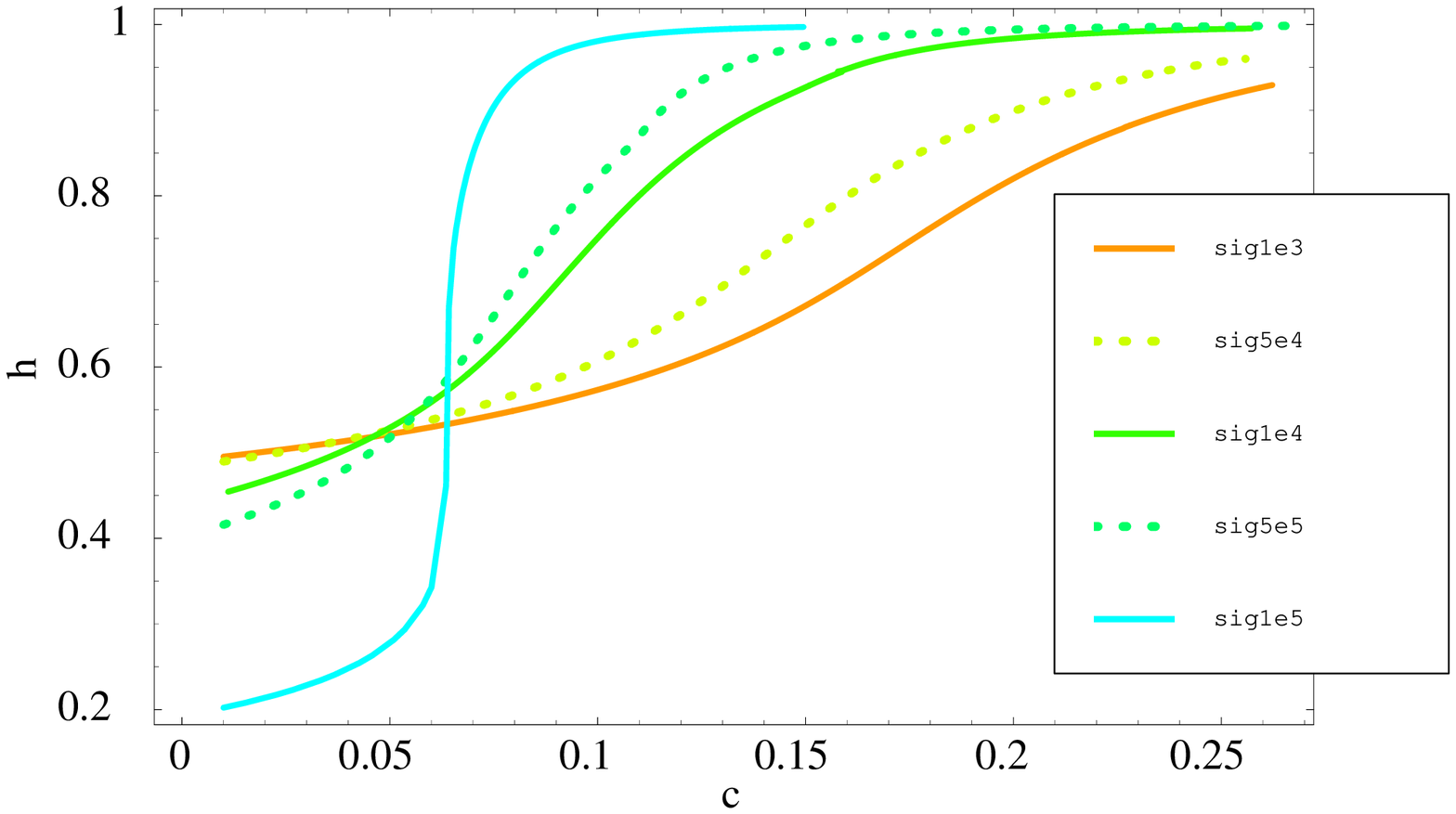}
\caption[$\Theta_{\rm R}$ versus $\chi$ for different $\sigma$]{The fraction of stiff segments is plotted as a function 
of $\chi$ for different values of $\sigma$ and $N=10^5$, $\epsilon=g=0$. The crossover becomes sharper with decreasing $\sigma$, 
i.e. increasing cooperativity.}
\label{fracH-sigma}
\end{center}
\end{figure}
Fig.(\ref{fracH-sigma}) shows that the crossover becomes sharper with increasing cooperativity (decreasing $\sigma$) as expected. 
For very high cooperativity ($\sigma = 10^{-5}$) and no selective interaction ($\chi \approx 0$), 
the fraction of stiff segments ($\Theta_{\rm R}=0.2$) is much smaller than $1/2$. When the transition point is approached, the fraction
of stiff segments increases rapidly. For lower cooperativity (larger $\sigma$), not only is the crossover much smoother but the 
fraction of stiff segments at $\chi \approx 0$ is much higher ($\Theta_{\rm R} \approx 0.5$ for $\sigma = 10^{-3}$ and 
$\sigma = 5\cdot10^{-4}$). Both features clearly show the cooperativity of the transition.

\begin{figure}[h!]
\begin{center}
\psfrag{m}{\large $S$}
\psfrag{c}{\large $-\chi$}
\psfrag{sig1e3}{\scriptsize $\sigma=10^{-3}$}
\psfrag{sig5e4}{\scriptsize $\sigma=5\cdot10^{-4}$}
\psfrag{sig1e4}{\scriptsize $\sigma=10^{-4}$}
\psfrag{sig5e5}{\scriptsize $\sigma=5\cdot10^{-5}$}
\psfrag{sig1e5}{\scriptsize $\sigma=10^{-5}$}
\includegraphics[width=0.85\linewidth]{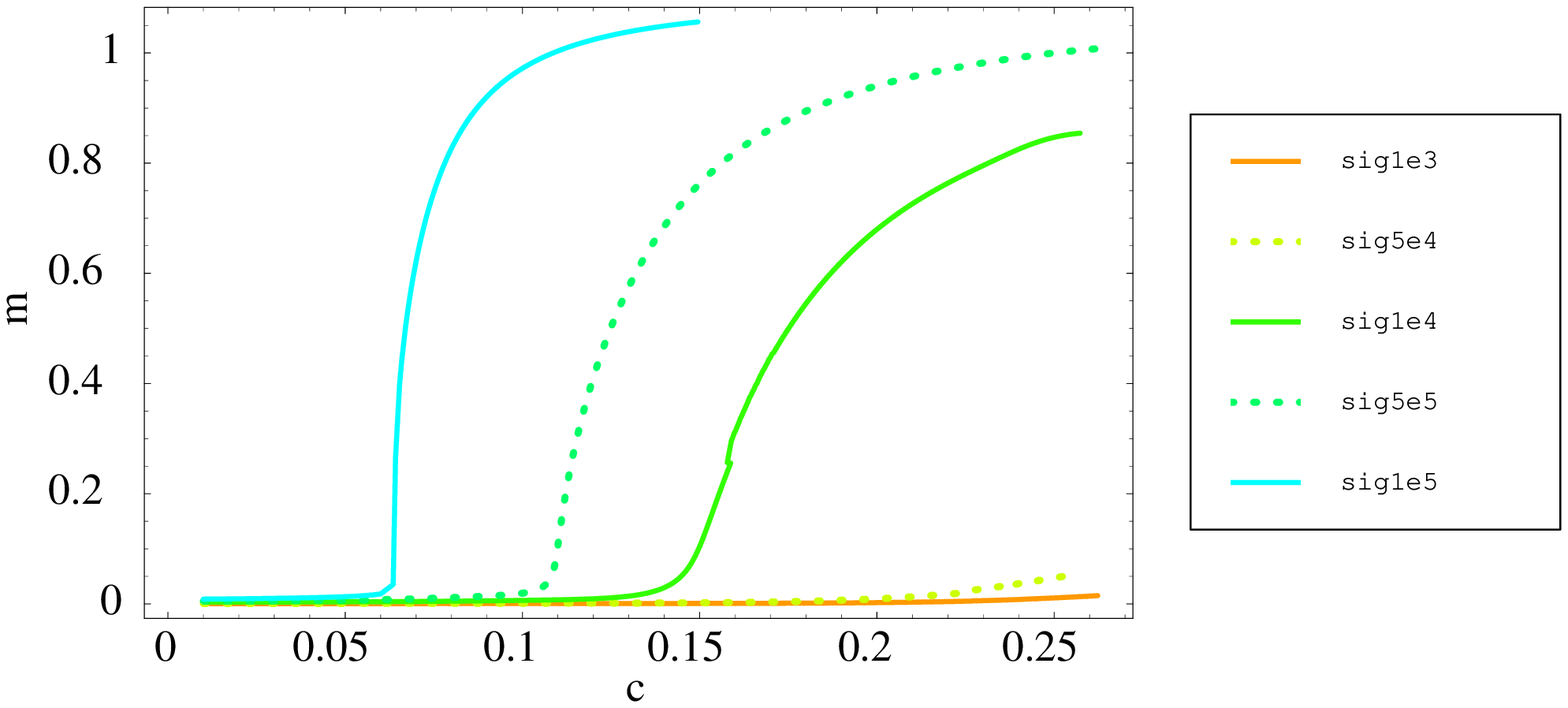}
\caption[$S$ versus $\chi$ for different $\sigma$]{The nematic order parameter $S$ is plotted as a function 
of $\chi$ for different values of $\sigma$ and $N=10^5$. The transition point is shifted to 
higher values of $|\chi|$ with increasing $\sigma$.}
\label{nem-sigma}
\end{center}
\end{figure}

The behavior of the nematic order parameter $S$ for different values of $\sigma$ is shown in Fig.(\ref{nem-sigma}). 
For higher $\sigma$ the increase in nematic order is less steep and the transition point is significantly shifted to 
higher values of $|\chi|$. For $\sigma=5\cdot10^{-4}$ and higher, the system only develops a slight onset of order 
and $S$ remains very small even deep in the globular state at $\chi=-0.25$. This again demonstrates the role of 
cooperativity in the formation of a liquid-crystalline globule.   

\begin{figure}[h!]
\begin{center}
\psfrag{L}{\large $L_{\rm R}$}
\psfrag{c}{\large $-\chi$}
\psfrag{sig1e3}{\scriptsize $\sigma=10^{-3}$}
\psfrag{sig5e4}{\scriptsize $\sigma=5\cdot10^{-4}$}
\psfrag{sig5e5}{\scriptsize $\sigma=5\cdot10^{-5}$}
\psfrag{sig1e5}{\scriptsize $\sigma=10^{-5}$}
\includegraphics[width=0.85\linewidth]{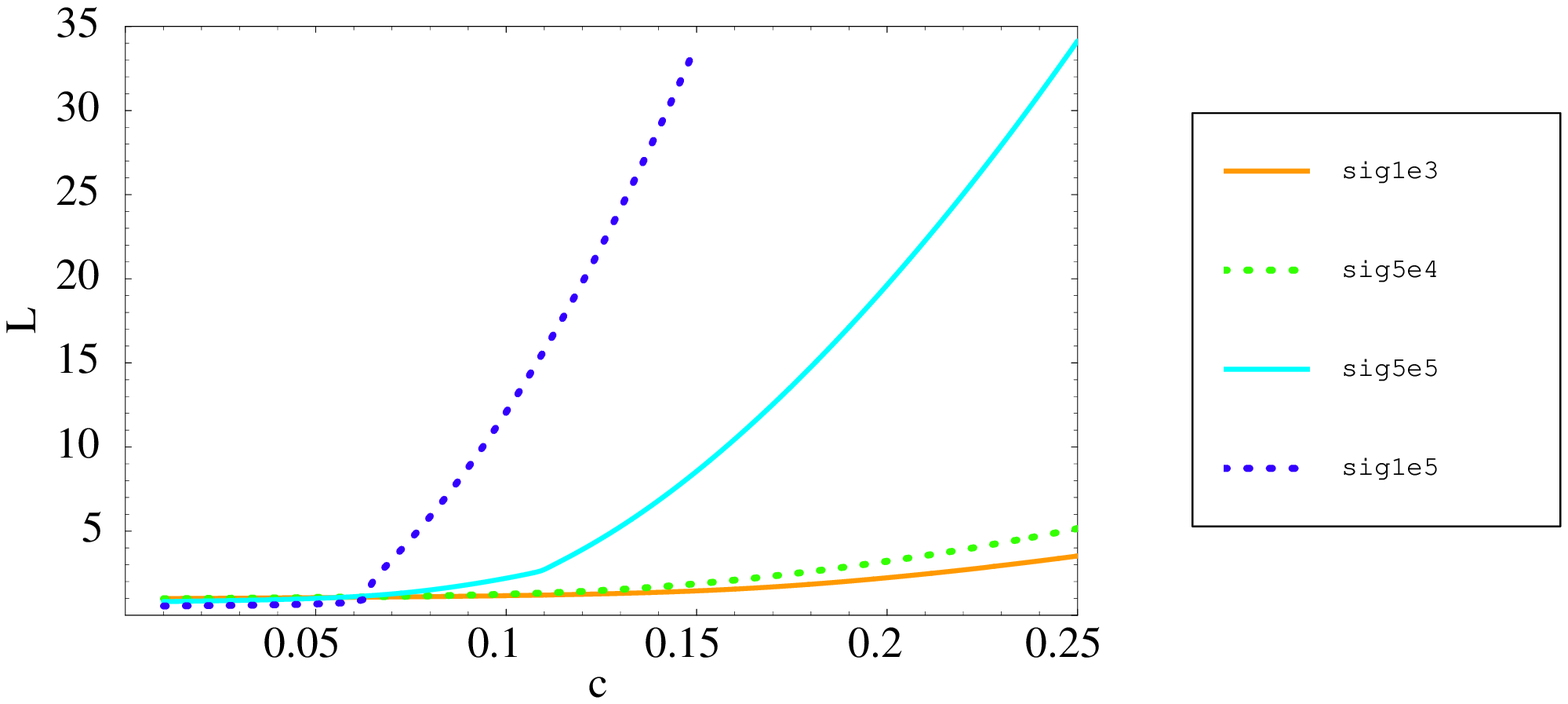}
\caption[$L_{\rm R}$ versus $\chi$ for different $\sigma$]{The average rod length $L_{\rm R}$ is plotted as a function 
of $\chi$ for different values of $\sigma$ and $N=10^5$.}
\label{Lh-sigma}
\end{center}
\end{figure}

It is therefore of interest to investigate the average rod length $L_{\rm R}$, which should also strongly depend on the 
cooperativity. 
Fig.(\ref{Lh-sigma}) shows that the increase of $L_{\rm R}$ with $|\chi|$ after the transition point becomes steeper for 
smaller values of $\sigma$ as one would expect.
The isotropic-nematic transition is enhanced by cooperativity and even becomes impossible if the 
cooperativity is too small. As can be seen from Fig.(\ref{Lh-sigma}), 
only a cooperative system forms long rods and not just many very short ones. Remember, that 
$\sigma$ is associated with the energy penalty for a boundary between rod and coil,  as discussed in the introduction . The smaller $\sigma$ is, the larger 
is this energy penalty and the more favorable is the formation of long rods rather than short ones. It is intuitively clear that 
long rods align much easier than very short ones. When the average rod length stays roughly equal to $1$ for high values of $|\chi|$, alignment 
cannot happen.
Hence cooperativity is important to drive this transition. 
\subsubsection{$\epsilon$-dependence}
In this subsection the $\epsilon$-dependence of the transition from amorphous to liquid-crystalline globule is investigated \cite{Nowak4}.
In Section IIB, it was shown that the fraction of stiff segments increases with increasing energy gain $\epsilon$ per stiff segment. 
A higher value of $\epsilon$ yields a higher offset of $\Theta_{\rm R}$ at $\chi=0$ which should in turn lead to a smaller value 
of $|\chi|$ at the transition point. 

\begin{figure}[h!]
\begin{center}
\psfrag{h}{\large $\Theta_{\rm R}$}
\psfrag{c}{\large $-\chi$}
\psfrag{eps00}{\small $\epsilon=0$}
\psfrag{eps0005}{\small $\epsilon=0.005$}
\psfrag{eps001}{\small $\epsilon=0.01$}
\psfrag{eps003}{\small $\epsilon=0.03$}
\psfrag{eps005}{\small $\epsilon=0.05$}
\psfrag{eps008}{\small $\epsilon=0.08$}
\psfrag{eps01}{\small $\epsilon=0.1$}
\psfrag{eps013}{\small $\epsilon=0.13$}
\psfrag{eps015}{\small $\epsilon=0.15$}
\psfrag{eps02}{\small $\epsilon=0.2$}
\psfrag{eps025}{\small $\epsilon=0.25$}
\includegraphics[width=0.9\linewidth]{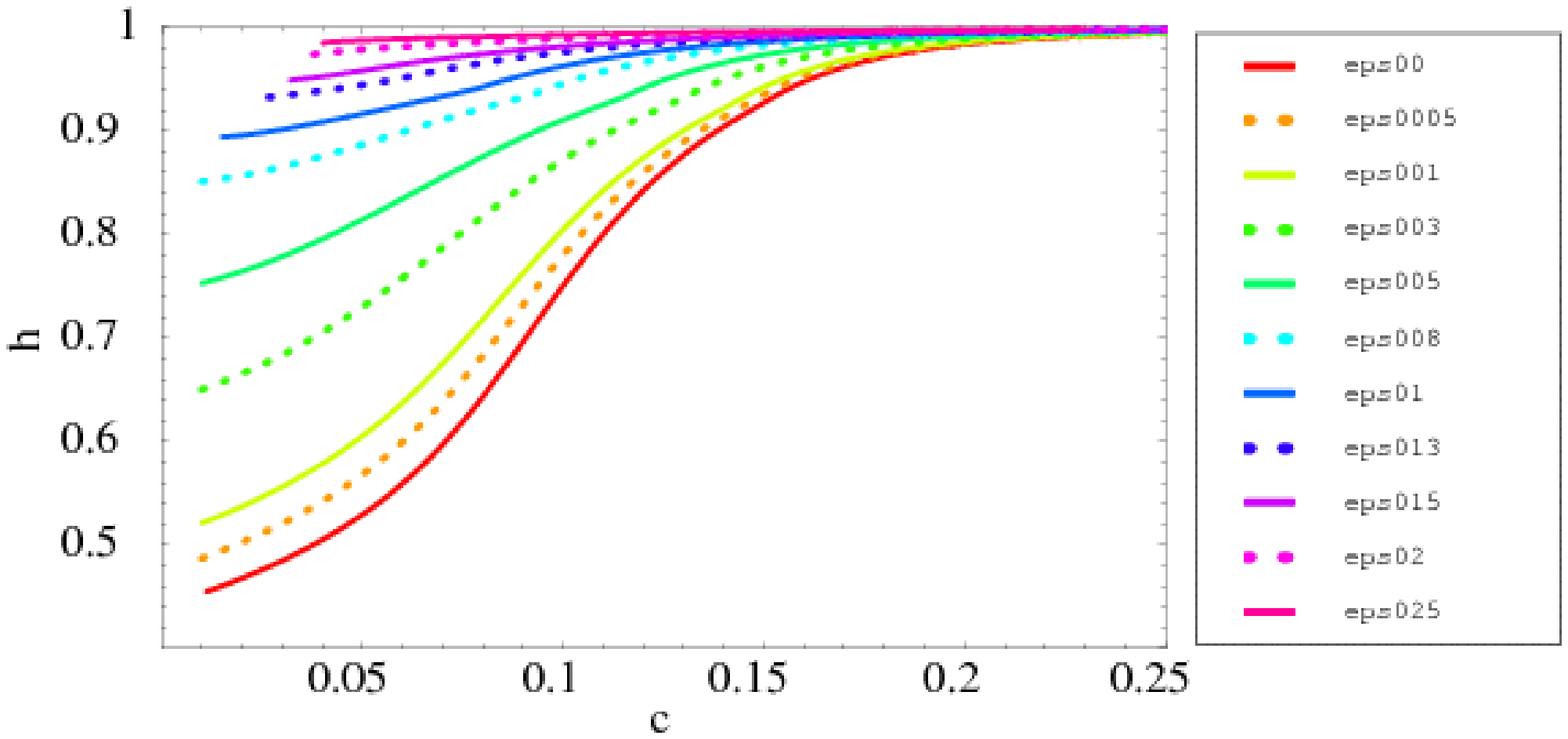}
\caption[$\Theta_{\rm R}$ versus $\chi$ for different $\epsilon$]{Fraction of helical segments $\Theta_{\rm R}$ as a function of $\chi$ 
for different values of the energy gain per helical segment $\epsilon$.} \label{fracH-epsilon}
\end{center}
\end{figure}

\begin{figure}[h!]
\begin{center}
\psfrag{m}{\large $S$}
\psfrag{c}{\large $-\chi$}
\psfrag{eps00}{\small $\epsilon=0$}
\psfrag{eps0005}{\small $\epsilon=0.005$}
\psfrag{eps001}{\small$\epsilon=0.01$}
\psfrag{eps003}{\small $\epsilon=0.03$}
\psfrag{eps005}{\small $\epsilon=0.05$}
\psfrag{eps008}{\small $\epsilon=0.08$}
\psfrag{eps01}{\small $\epsilon=0.1$}
\psfrag{eps013}{\small $\epsilon=0.13$}
\psfrag{eps015}{\small $\epsilon=0.15$}
\psfrag{eps02}{\small $\epsilon=0.2$}
\psfrag{eps025}{\small $\epsilon=0.25$}
\includegraphics[width=0.9\linewidth]{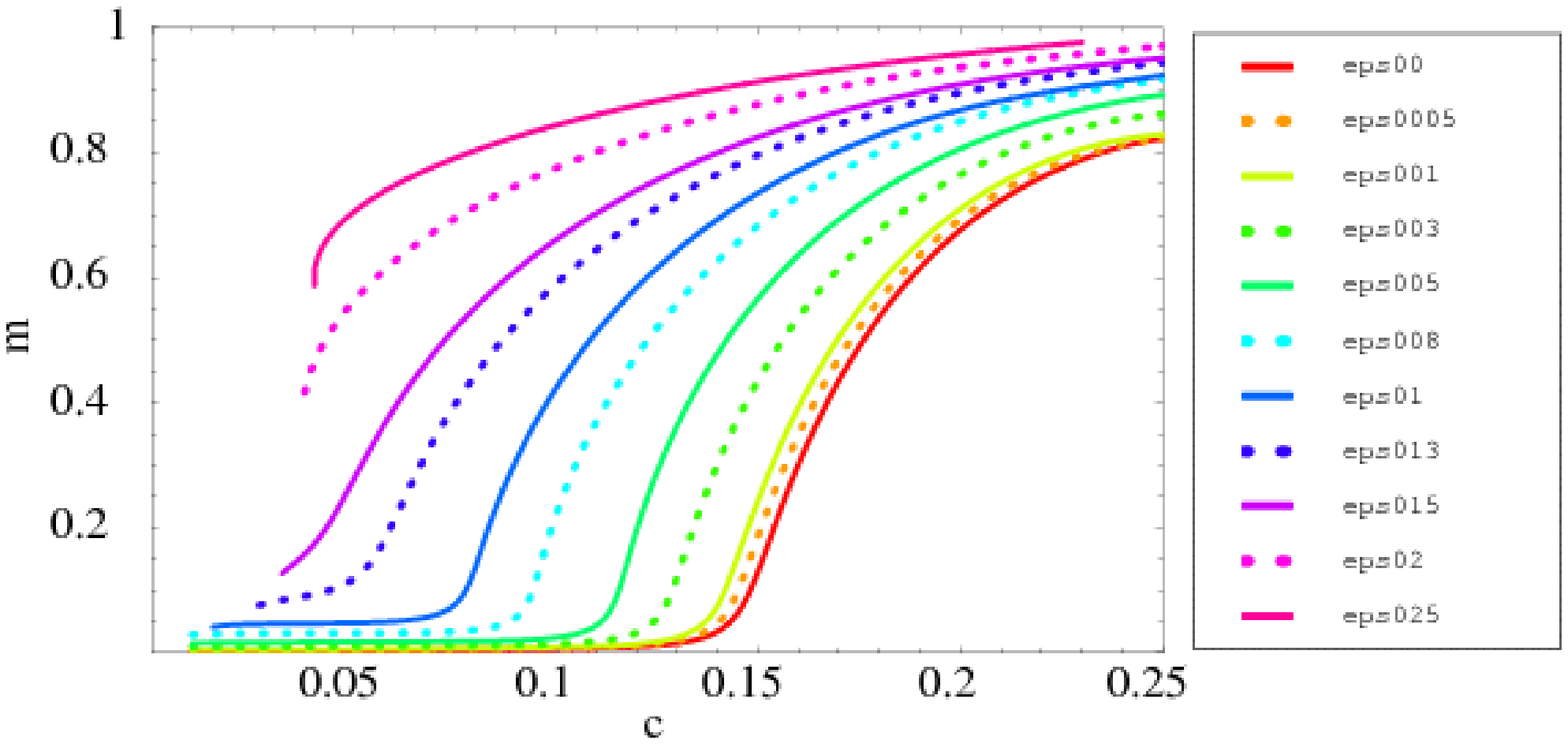}
\caption[$S$ versus $\chi$ for different $\epsilon$]{Nematic order parameter $S$ as a function of $\chi$ for different values of $\epsilon$.}
\label{nem-epsilon}
\end{center}
\end{figure}

Figs.(\ref{fracH-epsilon}, \ref{nem-epsilon}) demonstrate this behavior. For all plots in this 
subsection the cooperativity parameter is set to $\sigma=10^{-4}$ and the total chain length to $N=10^5$. The explicit rod-rod 
alignment interaction is still switched off ($g=0$). 
The onset of the transition is indeed shifted to lower values of $|\chi|$ with increasing energy gain 
per helical segments $\epsilon$. This is due to an increase of bulk interaction energy for fixed $\chi$ 
with increasing number of stiff segments. 

The curves for $\epsilon \ge 0.1$ in Figs.(\ref{fracH-epsilon}, \ref{nem-epsilon}) 
start at non-zero values of $\chi$. These values of $\chi$ correspond to the transition point between open chain regime
and globular regime, similar to the discussion in Section IIA. The main difference in this case is that at $v=-0.2$ the 
non-selective attractive two-body interaction is not strong enough to  drive the system into the globular 
state. The additional selective two-body interaction between the stiff segments only (with interaction parameter $\chi$) is needed to 
drive the system into the globular state.
That this is only the case for $\epsilon \ge 0.1$ 
can be explained as follows (see also Section IIB). If there is no additional selective interaction 
energy which favors a compactification of the stiff segments, the stiffening of parts of the chain due to an increase of 
$\Theta_{\rm R}$ with increasing $\epsilon$ pushes the chain segments further apart from each other and therefore leads to a more 
open structure. For $\epsilon \ge 0.1$, the system is thus pushed into the open chain regime at $\chi=0$. On the other hand, for $\epsilon \ge 0.17$ the system can be driven directly from open chain to liquid-crystalline globule as soon as $ |\chi| $ exceeds some crossover value.

\begin{figure}[h!]
\begin{center}
\psfrag{v}{\large $-\chi$}
\psfrag{z}{\large $R_z$}
\psfrag{r}{\large $R_\varrho$}
\includegraphics[width=0.49\linewidth]{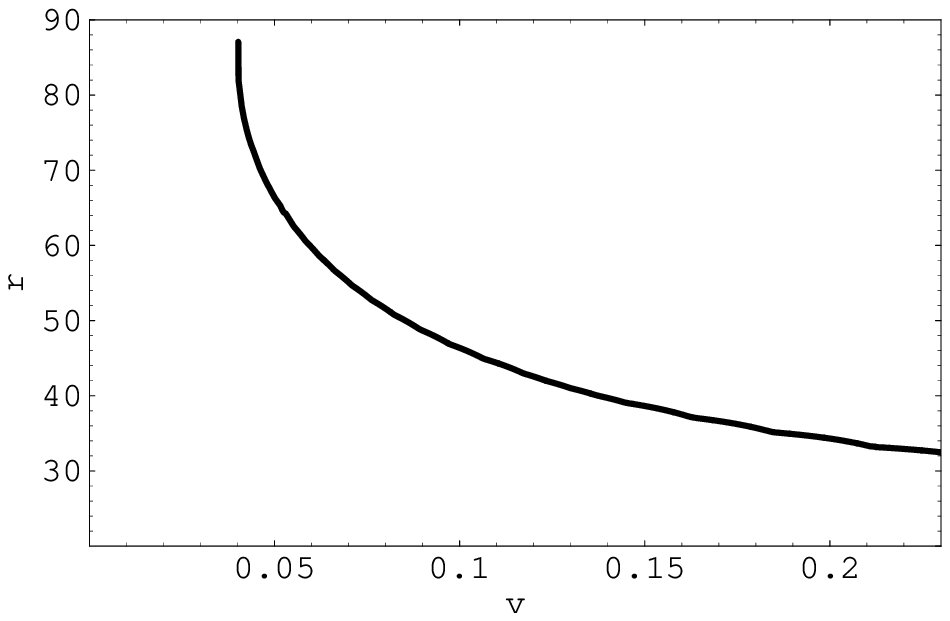}
\includegraphics[width=0.49\linewidth]{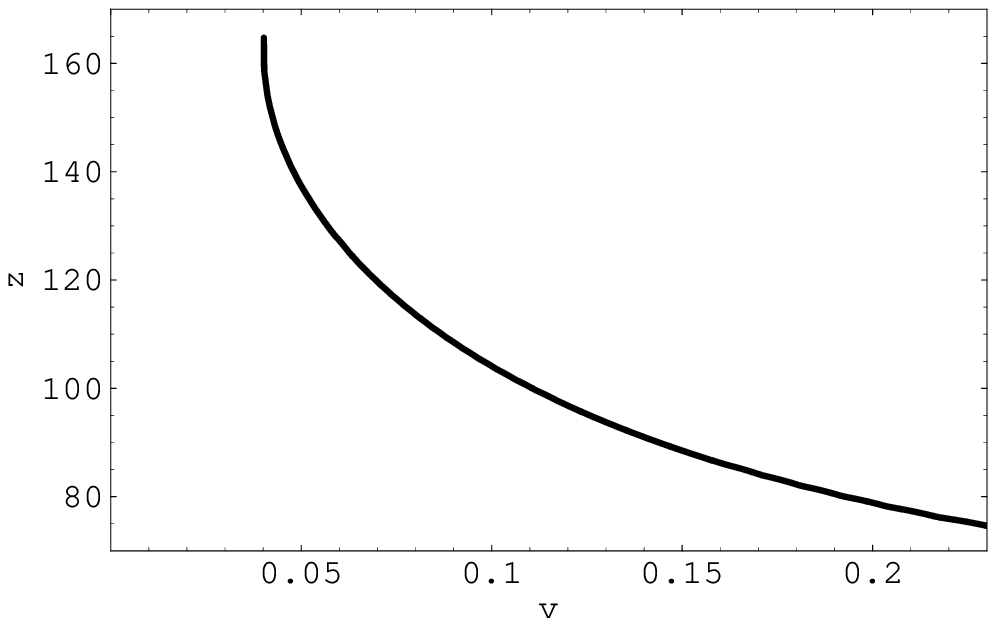}
\caption[$R_\varrho$ and $R_z$ versus $\chi$ for $\epsilon=0.25$]{The extensions of the globule in $\varrho$-direction $R_\varrho$ 
and in $z$-direction $R_z$ are plotted as functions of $\chi$ for $\epsilon=0.25$.}
\label{rho-z-coil-globule-eps25}
\end{center}
\end{figure}

\begin{figure}[h!]
\begin{center}
\includegraphics[width=0.8\linewidth]{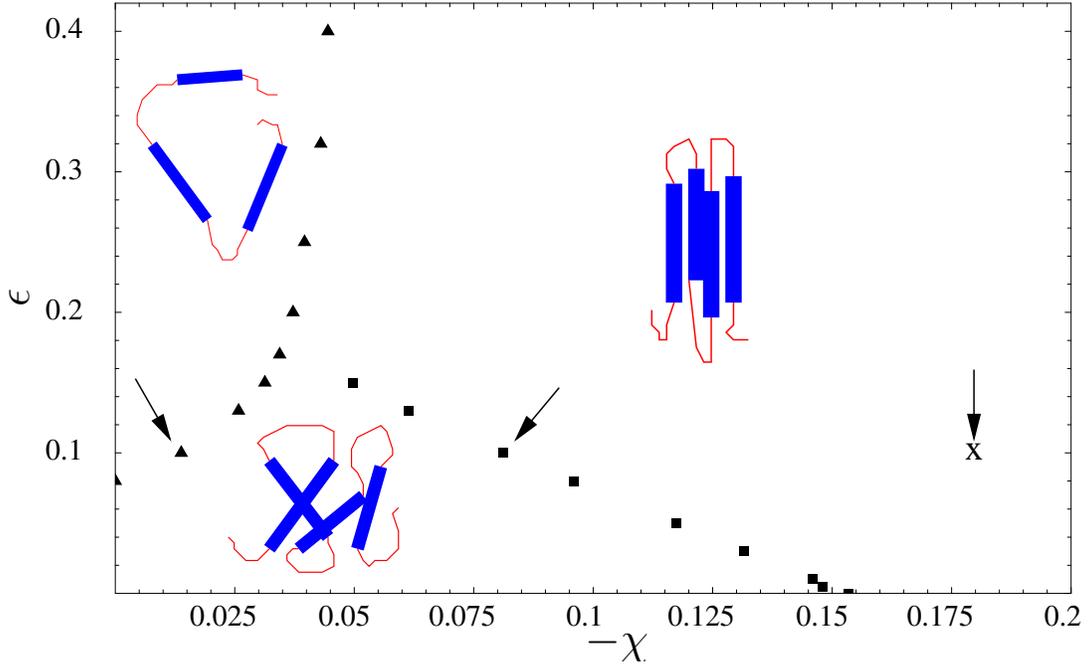}
\caption[Phase diagram of a rod-coil copolymer in $\epsilon$-$\chi$ space]{Phase diagram of a rod-coil copolymer in $\epsilon$-$\chi$ space. 
The upper left area corresponds to an open chain, 
the lower left area to an amorphous globule and the right area to a nematic liquid-crystalline-globule.
The little arrow to the left indicates the point in the phase diagram which corresponds to the top two 
pictures in Fig.(\ref{dens}). The arrow in the middle corresponds to the middle 
two pictures and the arrow to the right to the bottom ones. Parameters: $N=10^5$, $\sigma=10^{-4}$, $g=0$.}
\label{phasediagram}
\end{center}
\end{figure}

These findings permit the  
computation of a complete phase diagram of the rod-coil copolymer in $\epsilon$-$\chi$ space, see Fig.(\ref{phasediagram}). 
But before this is done, it has to be checked whether the definition of the transition point between coil and globule as the minimum 
of the $N(\mu)$-curve is also reasonable in the case of strong nematic order. 
In Section IIA  the radius $R$ of the globule (defined as the point at which the total density has 
decreased to $\rho(R) = 10^{-3}\rho_0$) was plotted as a function of $v$, see Fig.(\ref{v-R-coil-globule}). 
It showed a rapid increase when the transition point between 
globule and coil was approached. Here the globule shows nematic order and has an asymmetric shape, i.e. it is   necessary to distinguish 
between $\varrho$- and $z$-direction. The extensions of the globule in $\varrho$-direction $R_\varrho$ and in $z$-direction $R_z$ are 
defined - analogously to $R$ in the case of a spherical globule - as $\rho(R_\varrho,0) = \rho(0,R_z) =10^{-3}\rho_0$.
In Fig.(\ref{rho-z-coil-globule-eps25}) $R_\varrho$ and 
$R_z$ are plotted as functions of $\chi$ for $\epsilon=0.25$. For this choice of $\epsilon$, the system already shows strong nematic order 
at $\chi=-0.0396$ which corresponds to the transition point between globule and coil.
Fig.(\ref{rho-z-coil-globule-eps25}) demonstrates that the definition of the transition point between open chain and globular regime as the minimum 
of the $N$-$\mu$ curve is valid even if the system shows strong nematic order. Both curves in Fig.(\ref{rho-z-coil-globule-eps25}) 
show a rapid increase when approaching the transition point from the right. The different values of $R_\varrho$  and $R_z$ reflect the 
cylindrical shape of the globule due to the nematic order of the stiff segments. For $\epsilon=0.25$ no isotropic globular phase exists 
and the crossover leads directly to a liquid-crystalline globule. 
  
\begin{figure}[h!]
\begin{center}
\psfrag{v}{\large $-\chi$}
\psfrag{z}{\large $R_z$}
\psfrag{r}{\large $R_\varrho$}
\includegraphics[width=0.49\linewidth]{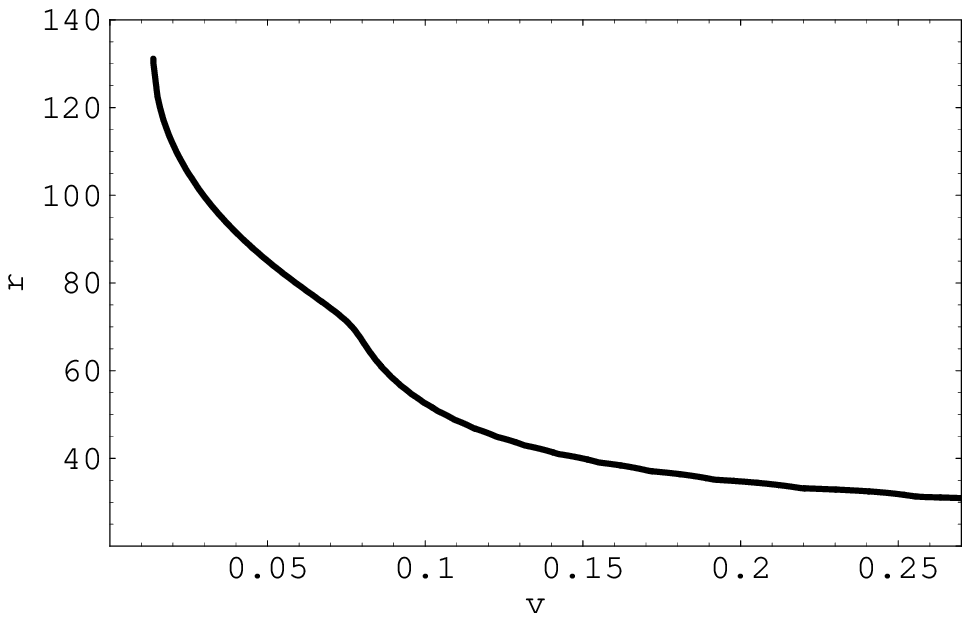}\includegraphics[width=0.49\linewidth]{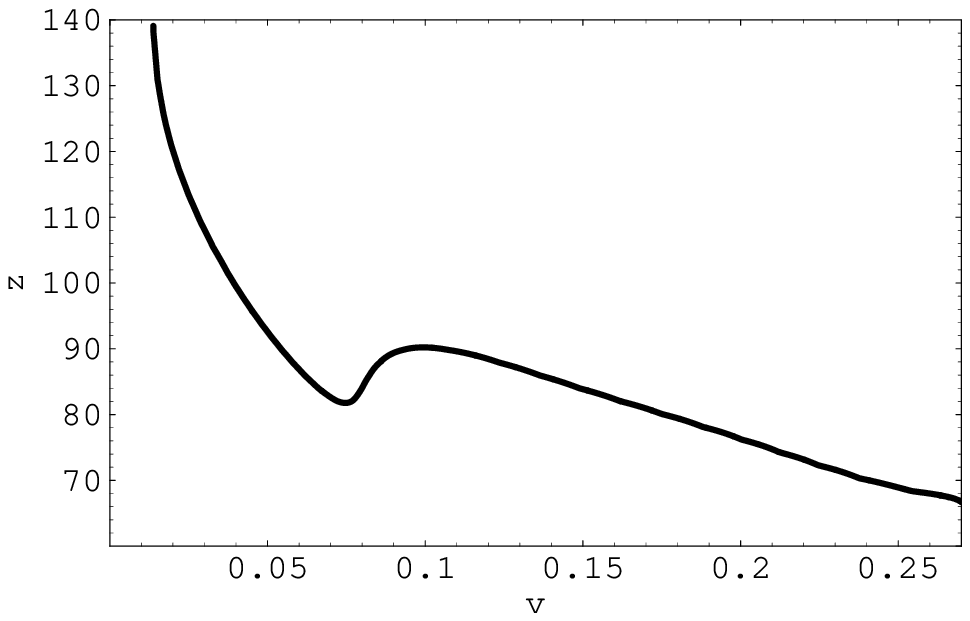}\caption[$R_\varrho$ and $R_z$ versus $\chi$ for $\epsilon=0.1$]{The extensions of the globule in $\varrho$-direction $R_\varrho$ 
and in $z$-direction $R_z$ are plotted as functions of $\chi$ for $\epsilon=0.1$.}
\label{rho-z-coil-globule-eps1}
\end{center}
\end{figure}

For $\epsilon=0.1$ on the other hand, the system shows no nematic order at the transition point between open chain and globule ($\chi=-0.0138$).
The system therefore undergoes two transitions. First from an open chain to an amorphous globule and then at higher $|\chi|$ from an amorphous to a liquid-crystalline globule. In Fig.(\ref{rho-z-coil-globule-eps1})
$R_\varrho$ and $R_z$ are plotted as functions of $\chi$. 
When the transition point between open chain and globule (at $\chi=-0.0138$) is approached from the right, $R_\varrho$ and $R_z$ show the expected strong 
increase. When the transition point between amorphous and liquid-crystalline globule (at $\chi=-0.0812$) is approached from the left, 
$R_\varrho$ and $R_z$ show a different behavior. In a small interval (corresponding to the width of the transition)
$R_z$ increases whilst $R_\varrho$ decreases even stronger than before. In this interval the asymmetric shape of the globule is developed. 
This behavior will be discussed further in Subsection D.   

As already mentioned above, it is now possible to compute a complete phase diagram in $\epsilon$-$\chi$ space. This  phase diagram is shown 
in Fig.(\ref{phasediagram}).
The triangles are the transition points between open chain and globule. The squares are the transition points 
between amorphous and liquid-crystalline globule. 
Note, that the points plotted in the phase diagram, Fig.(\ref{phasediagram}), are what is defined 
above as the points of rather broad crossover regions. Therefore the boundaries in the 
phase diagram have to be understood as center lines of broader regions in which the crossover from 
one phase to the other occurs. The little arrows in Fig.(\ref{phasediagram}) correspond each to one of the rows of the color coded density plots 
shown in Fig.(\ref{dens}), helping to illustrate how the density profiles look at these points in the phase diagram. 
\subsubsection{$g$-dependence}
The transition from amorphous to liquid-crystalline globule occurs without an explicit angle-dependent alignment interaction between the rods. 
It is nevertheless interesting to switch the explicit rod-rod alignment interaction on and study its influence 
on the transition. The alignment interaction is of the Maier-Saupe form  and its strength is controlled 
by the interaction parameter $g$, see Eq.(\ref{F-exp}). It is chosen to be attractive to favor alignment of the rods. 

\begin{figure}[h!]
\begin{center}
\psfrag{h}{\large $\Theta_{\rm R}$}
\psfrag{c}{\large $-\chi$}
\psfrag{g0}{\small $g=0$}
\psfrag{g05}{\small $g=-0.05$}
\psfrag{g1}{\small $g=-0.1$}
\psfrag{g15}{\small $g=-0.15$}
\psfrag{g2}{\small $g=-0.2$}
\includegraphics[width=0.96\linewidth]{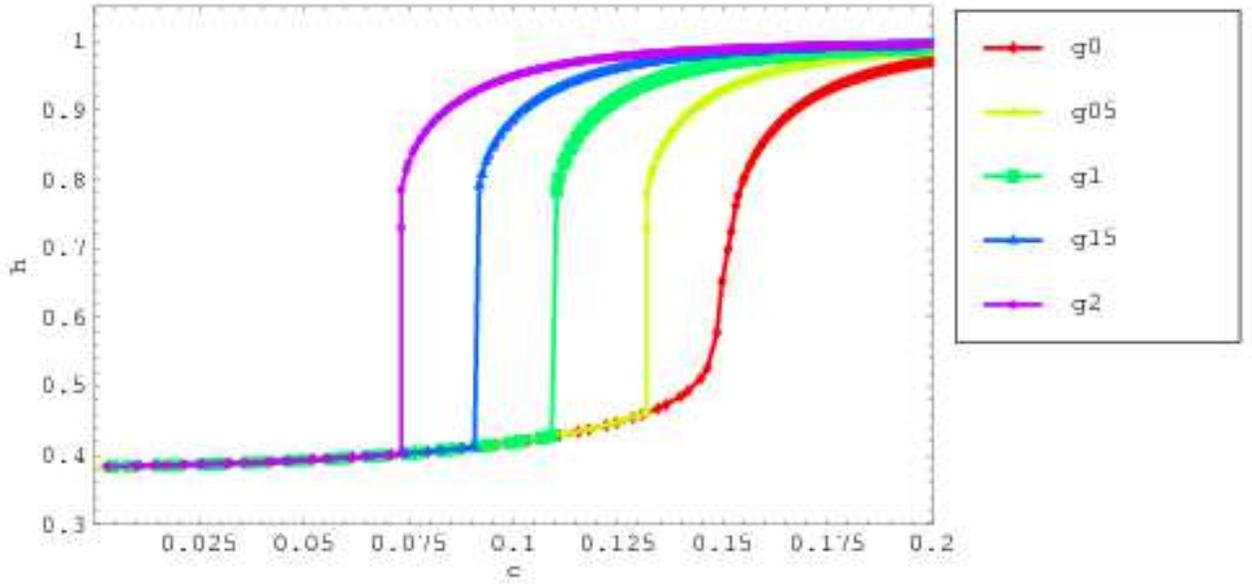}
\caption[$\Theta_{\rm R}$ versus $\chi$ for different $g$]{Fraction of helical segments $\Theta_{\rm R}$ as a function of $\chi$ 
for different values of $g$.} 
\label{fracH-g}
\end{center}
\end{figure}

\begin{figure}[h!]
\begin{center}
\psfrag{h}{\large $S$}
\psfrag{c}{\large $-\chi$}
\psfrag{g0}{\small $g=0$}
\psfrag{g05}{\small $g=-0.05$}
\psfrag{g1}{\small $g=-0.1$}
\psfrag{g15}{\small $g=-0.15$}
\psfrag{g2}{\small $g=-0.2$}
\includegraphics[width=0.96\linewidth]{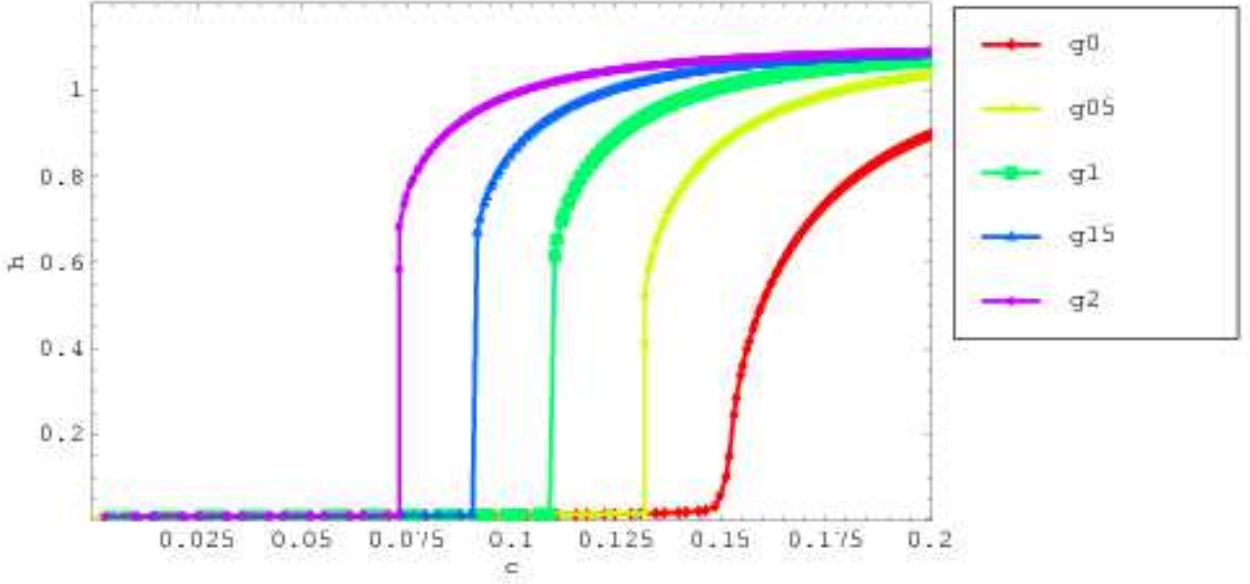}
\caption[$S$ versus $\chi$ for different $g$]{Nematic order parameter $S$ as a function of $\chi$ for different values of $g$.}
\label{nem-g}
\end{center}
\end{figure}

In Fig.(\ref{fracH-g}) 
the fraction of stiff segments is plotted as a function of $\chi$ for different values of $g$ and $\sigma=10^{-4}$, $\epsilon=0$, $N=9.5\cdot10^3$. 
The corresponding nematic order parameter $S$ is shown in Fig.(\ref{nem-g}). 
The transition is shifted to lower values of values of $|\chi|$ with increasing $|g|$. This is not suprising, since for $g<0$ the 
alignment term provides an additional incentive for the system to generate nematic order and stiff segments. But the fact that 
almost up to the respective transition points the $\Theta_{\rm R}(\chi)$-curves lie perfectly on top of each other is quite remarkable. 
This can be explained as follows. For $\psi_2\equiv 0$ the alignment term in Eq.(\ref{F-exp}) is equal to zero. Therefore the system 
has to generate a finite $\psi_2$ (i.e. at least small nematic order) before the alignment interaction can have an effect on the system.
As long as there is no nematic order in the system ($\psi_2\equiv 0$) the alignment term is zero and the system behaves as the one 
with $g=0$.
As expected, the attractive alignment interaction enhances the generation of nematic order and thus shifts the transition point to lower values of 
$|\chi|$.
\subsection{Free energy}
\begin{figure}[h!]
\begin{center}
\psfrag{c}{\large $-\chi$}
\psfrag{F}{\large $F$}
\includegraphics[width=0.7\linewidth]{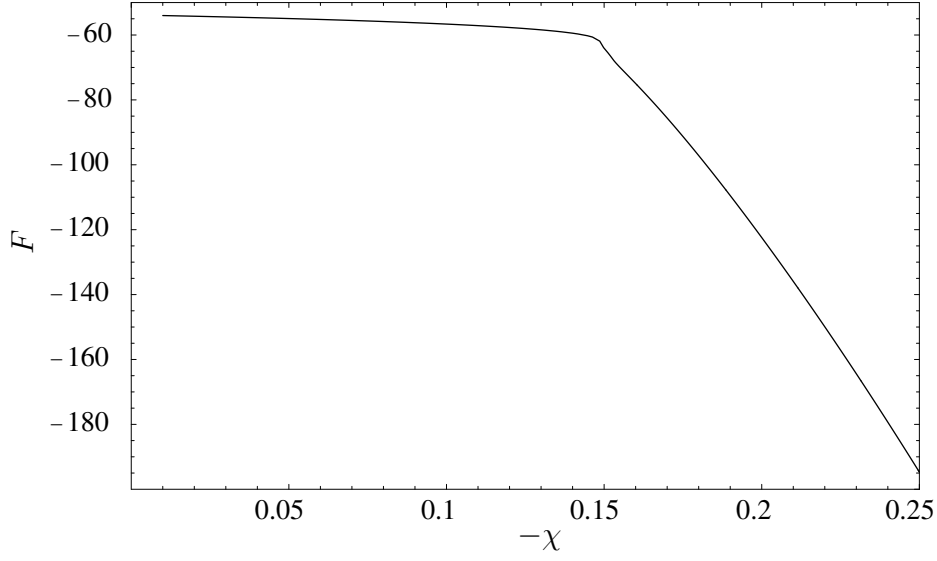}
\caption[Plot of $F$ versus $\chi$]{Free energy $F$ plotted as a function of $\chi$ for $\sigma=10^{-4}$, $N=9.5\cdot10^3$ and $\epsilon = g = 0$.}
\label{F-plot}
\end{center}
\end{figure}

\begin{figure}[h!]
\begin{center}
\psfrag{c}{\large $-\chi$}
\psfrag{F}{\large $F$}
\psfrag{FgradR}{\small $F^{\rm surf}_\varrho$}
\psfrag{FgradZ}{\small $F^{\rm surf}_z$}
\includegraphics[width=0.65\linewidth]{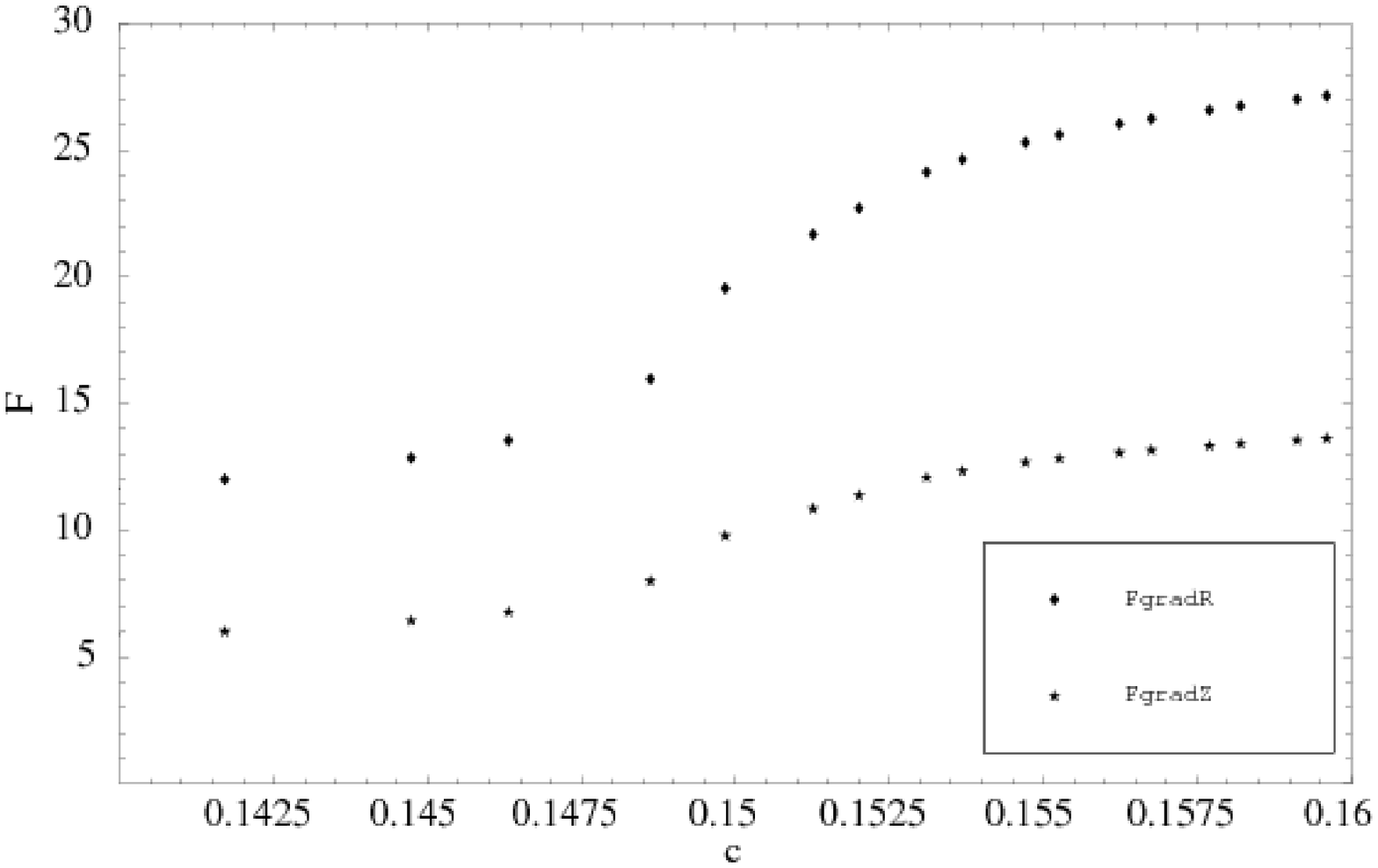}
\caption[Plot of $F^{\rm surf}_\varrho$ and $F^{\rm surf}_z$ versus $\chi$]{$F^{\rm surf}_\varrho$ and $F^{\rm surf}_z$ plotted as 
functions of $\chi$ for $\sigma=10^{-4}$, $N=9.5\cdot10^3$ and $\epsilon = g = 0$.}
\label{F-plot-grad}
\end{center}
\end{figure}

\begin{figure}[h!]
\begin{center}
\psfrag{c}{\small $-\chi$}
\psfrag{F}{\small $F/A$}
\psfrag{L}{\small $R$}
\psfrag{Rlim}{\scriptsize $R_{\varrho}$}
\psfrag{Zlim}{\scriptsize $R_z$}
\psfrag{FgradRnom}{\scriptsize $F^{\rm surf}_\varrho/(R_{\varrho}R_z)$}
\psfrag{FgradZnom}{\scriptsize $F^{\rm surf}_z/R_{\varrho}^2$}
\includegraphics[width=0.49\linewidth]{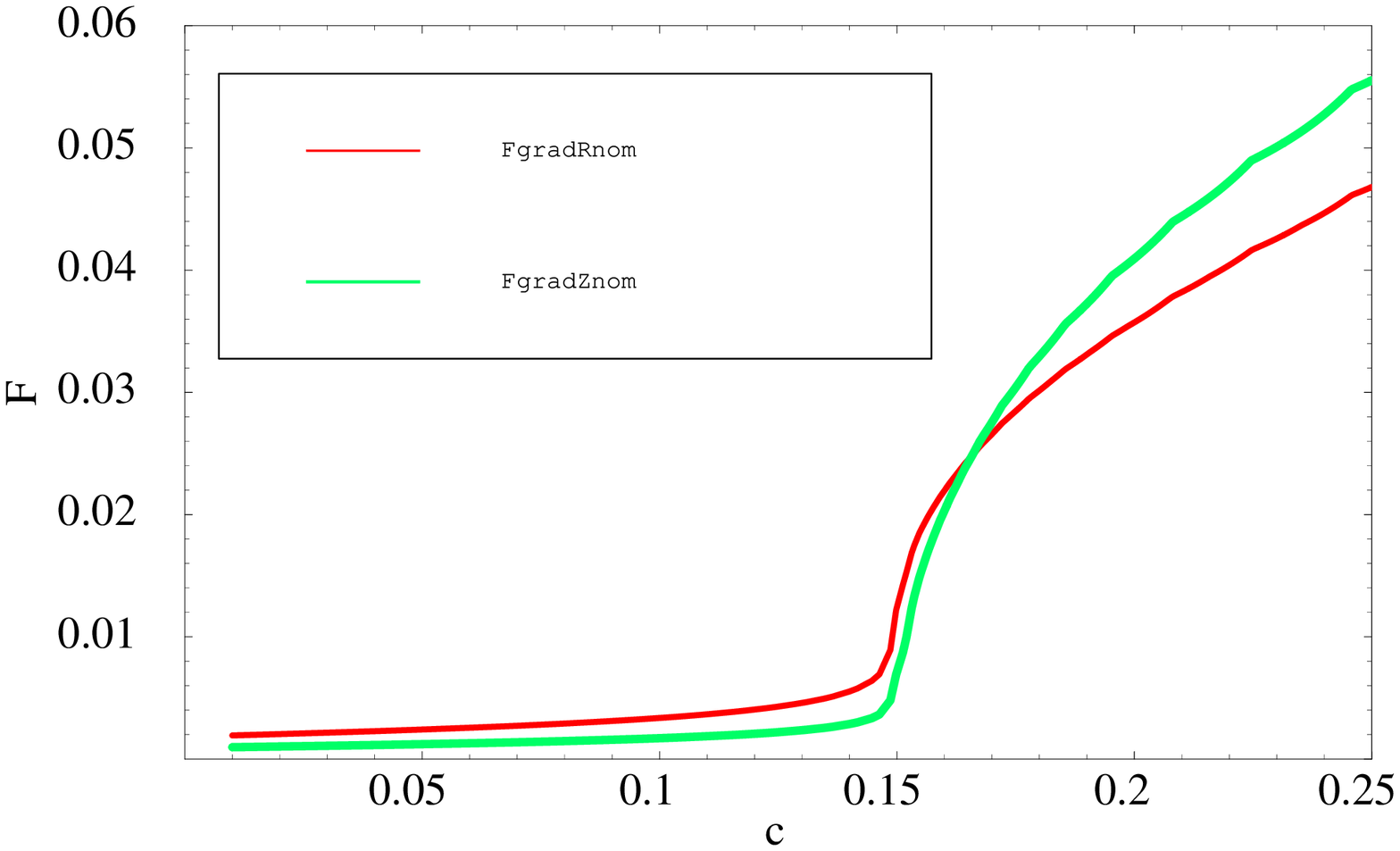}
\includegraphics[width=0.49\linewidth]{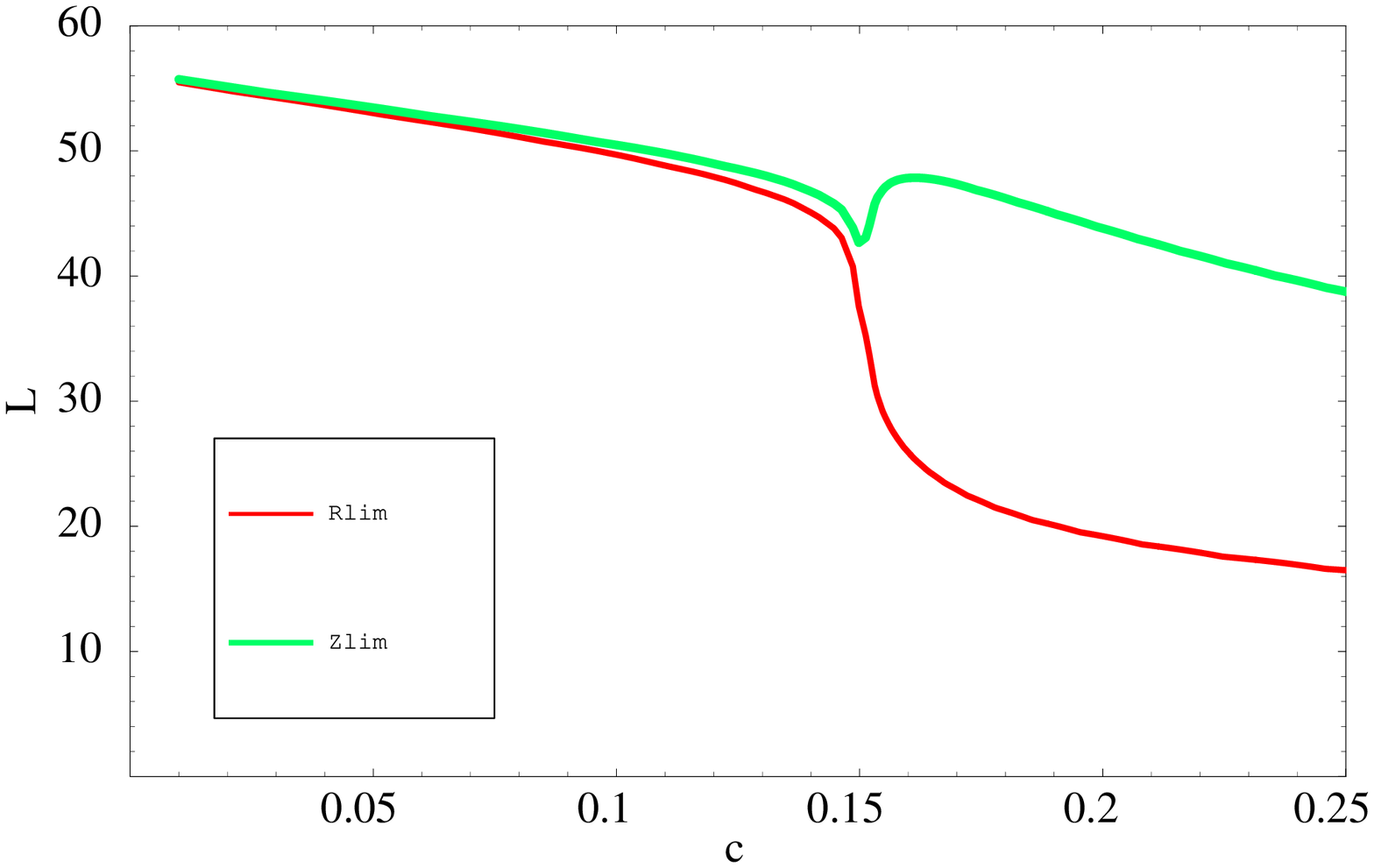}
\caption[Plot of surface energy per area and global extensions versus $\chi$]{$F^{\rm surf}_\varrho/(R_{\varrho}R_z)$ and 
$F^{\rm surf}_z/(R_{\varrho}^2)$ plotted as 
functions of $\chi$ for $\sigma=10^{-4}$, $N=9.5\cdot10^3$ and $\epsilon = g = 0$.}
\label{F-plot-grad-norm}
\end{center}
\end{figure}

In this subsection the behavior of the effective free energy in the crossover region is 
investigated. 
The effective saddle point grand potential $\Omega$ is given by Eq.(\ref{F-exp}). With the definition of 
the chemical potential used here the corresponding 
effective saddle point free energy is given by 
\begin{eqnarray}
F = \Omega - \mu N.
\end{eqnarray}
In Fig.(\ref{F-plot}) the free energy $F$ is plotted as a function of $\chi$ for $\sigma=10^{-4}$, $N=9.5\cdot10^3$ and $\epsilon = g = 0$. 
These parameters are the same as the ones used in Fig.(\ref{iso-nem1}).
Fig.(\ref{F-plot}) demonstrates that the free energy is changing its slope in the crossover region around the transition point as 
would be expected for a crossover. 
At the beginning of Section IIC the occurrence of the transition from amorphous to liquid-crystalline globule 
was explained in terms of the interplay between 
surface energy and bulk interaction energy. Because of the  anisotropy of the surface contributions which originate from the entropy of the rods, the globule tries to minimise its surface in $z$-direction and to maximise it in $\varrho$-direction. Before the transition 
the amorphous globule has a spherical shape. The surface energies in $x$-, $y$- and $z$-direction should therefore be all the same. For 
the cylindrical coordinates used here that implies $F^{\rm surf}_\varrho = 2F^{\rm surf}_z$. Fig.(\ref{F-plot-grad}) shows that this is indeed the case. 
But also in the liquid-crystalline globule regime after the transition their ratio is roughly equal to 2 as can be seen from Fig.(\ref{F-plot-grad}).
The extension of the globule in $\varrho$- and $z$-direction on the other hand is now very different. It is therefore instructive to plot 
$F^{\rm surf}_\varrho$ and $F^{\rm surf}_z$ normalized by the corresponding cross sections of the globule, which should give a feeling how the surface tension behaves. As an approximation of 
the cross section in $z$-direction $R_{\varrho}^2$ is chosen. The cross section in $\varrho$-direction is approximated by 
$R_{\varrho}R_z$, where $R_{\varrho}$ and $R_z$ are defined as in Subsection IIC3. In Fig.(\ref{F-plot-grad-norm}) the normalized 
surface contributions are plotted on the left hand side and $R_{\varrho}$ and $R_z$ one the right hand side.
The plot on the left shows that the surface energy per area in $z$-direction becomes indeed larger than the one in $\varrho$-direction in the 
interval in which the crossover from amorphous globule to liquid-crystalline globule occurs. The corresponding extensions of the globule 
$R_{\varrho}$ and $R_z$ plotted on the right illustrate the enlargement of the globule in $z$-direction and the diminution in $\varrho$-direction in the crossover interval. When the final shape of the liquid-crystalline globule has developed 
the curves become parallel again and decrease both indicating the further compactification of the entire globule.
The investigations of the surface contributions to the free energy further visualize that it is the anisotropy of the entropic surface energy 
that drives the system into the nematic state. 
\section{Conclusions}
The numerical solutions of the self-consistent field equations show that the rod-coil copolymer with variable composition can form 
three phase states, open chain, amorphous globule and nematic liquid-crystalline globule with high fraction of stiff segments.
The transition between the first two states is similar to the coil-globule transition of a homopolymer. The formation of a liquid-crystalline globular 
state without explicit alignment interaction between the rods is a novel result and deserves further discussion.
 
The formation of a liquid-crystalline globule 
from a rod-coil multiblock copolymer with fixed composition has been discussed for the first time in an early theoretical work~\cite{grosberg1}.  Phase states similar to the ones 
summarized in the phase diagram in Fig.(\ref{phasediagram}) have  been seen in~\cite{pitard}, where the authors 
consider a homopolymer in which each monomer carries a dipole moment and take into account explicit dipole-dipole interactions. 
However, the schematic phase diagram in~\cite{pitard} has been considered within the so-called volume approximation where the 
contribution of the surface energy can be neglected (at $N \rightarrow \infty$)~\cite{grosberg0} and the transition into 
the anisotropic globular state with nematic order is driven by dipole-dipole interactions. 
 
Recently, in the paper by Marenduzzo et al.~\cite{maritan} a new class of models for chain molecules has been considered . This model can be  viewed as an elastic tube and the concept of the chain thickness ("thick polymer") has been introduced by means of a specially prescribed  three-body interaction potential. Nevertheless, we should stress that on the mean-field level the consideration (within the so-called "chain of coins" model) is still limited to the  {\it volume approximation},  and the explicit alignment interaction in form of an Onsager second virial term is present (see the Eq.(14) in ref. ~\cite{maritan}).  In contrast to this in our case the formation of a LC-globule occurs without explicit alignment interactions between the helical parts. It is the entropic surface tension anisotropy which drives the globule in the nematic LC-state.

In an early work~\cite{kim} a cooperative helix-coil liquid-crystal transition has been found, very similar to the transition from amorphous  
to liquid-crystalline globule discussed here. The formation of nematic order is also accompanied by a strong increase in fraction of stiff 
(or helical) segments. The main difference is again that the transition is driven by an explicit alignment interaction of Maier-Saupe type
similar to the additional interaction ($g<0$) considered in Section IIC4 and not by the entropic surface tension anisotropy.

Monte Carlo simulations of a simple homopolymer model on the lattice have been carried out to study the interplay between coil-globule transition and $\alpha$-helix formation\cite{Shakhnovich}. The phase diagram has been presented in terms of two energetic parameters that characterize the hydrophobic attraction between monomers that are far apart in sequence and the local helix-stabilizing interaction. This is reminiscent of our $\chi - \epsilon$ diagram given in Fig.\ref{phasediagram}. The number of helical units, $\left\langle \Xi\right\rangle $,  (helicity) and the number of pairwise contacts, $\left\langle NC\right\rangle $,   are used  as  criteria  to distinguish among different phases.  Following these criteria the authors are  able to see four different phases.  Nevertheless it is difficult to compare these results with our findings because the order parameters we use are substantially different form the ones used in ref.\cite{Shakhnovich} .

In the context of secondary and tertiary structure formation in proteins the interplay of helix formation and liquid-crystalline order has been 
studied, see for instance~\cite{schulten,saven}. 
It was shown that liquid-crystalline ordering enhances the number of helical segments as well as the average 
length of the helices which is qualitatively similar to the results of the model presented here. This indicates that the model of a rod-coil 
multiblock copolymer with variable composition might be a good candidate to give a simple explanation for the formation of helix bundles in certain globular proteins. Both, simulations~\cite{Zhou} and experiments~\cite{Ptitsyn} show that proteins can adopt not only the native and completely denatured state (open chain) but also so-called  premolten and molten globular states. For helix-bundle proteins the premolten globule, which does not show any order of the helices, corresponds to the amorphous globule in our model. The molten globule with ordered helices but without native contacts (and therefore also without the characteristic helix-helix angle found in the native state) corresponds to the liquid-crystalline globule.   
Our model also shows that during the transition to a liquid-crystalline globule not only the amount of helical segments increases strongly 
but the globule also becomes more compact. The experimentally observed~\cite{uversky}  correlation between the amount of secondary structure elements 
and compactness of proteins mentioned in the introduction might therefore also be explained by this transition, at least in the case 
of helix bundle proteins.  

Irrespective of the possible application to helix bundle formation in proteins, the model which we have discussed in this paper  provides a relatively simple example of the general 
interplay between secondary structure (helices or stiff rods) and tertiary structure (liquid-crystalline order) in homopolypeptides.

\end{document}